\documentclass[aps,notitlepage,superscriptaddress,pra,twocolumn,nofootinbib,floatfix]{revtex4-1}

\usepackage[T1]{fontenc}
\usepackage{lmodern}
\usepackage{graphicx}
\usepackage{braket}
\usepackage{subfigure}
\usepackage{amsmath,amsfonts,mathrsfs,bbm,bm}
\usepackage{hyperref}
\hypersetup{
    colorlinks=true,       
    linkcolor=cyan,        
    citecolor=magenta,     
    filecolor=cyan,        
    urlcolor=cyan,         
    runcolor=cyan
    }
\usepackage[table,dvipsnames]{xcolor}
\newcolumntype{g}{>{\columncolor{gray!20}}r}
\usepackage[capitalize]{cleveref}

\input{revtexFix.tex}

\newcommand{\ii}{{i\mkern1mu}} 
\newcommand{\sigx}{\sigma^{x}}
\newcommand{\sigy}{\sigma^{y}}
\newcommand{\sigz}{\sigma^{z}}
\newcommand{\Phibar}{\bar{\Phi}}
\newcommand{\Psibar}{\bar{\Psi}}
\newcommand{\phibar}{\bar{\phi}}
\newcommand{\psibar}{\bar{\psi}}
\newcommand{\etad}{\eta^{\dagger}}
\newcommand{\tildeG}{\tilde{G}}

\DeclareMathOperator*{\argmin}{argmin}
\DeclareMathOperator*{\argmax}{argmax}
\DeclareMathOperator{\Tr}{Tr}

\newcommand{\mytensor}[1]{\mathbf{#1}}

\date{\today}
\begin{document}

\title{Finite temperature quantum annealing solving exponentially small gap problem with
  non-monotonic success probability}

\author{Anurag Mishra}
\email[Corresponding author:\ ]{anuragmi@usc.edu}
\affiliation{Department of Physics and Astronomy, University of  Southern California, Los
  Angeles, California 90089, USA}
\affiliation{Center for Quantum Information Science \& Technology, University of Southern
California, Los Angeles, California 90089, USA}

\author{Tameem Albash}
\affiliation{Department of Physics and Astronomy, University of Southern California, Los
  Angeles, California 90089, USA}
\affiliation{Center for Quantum Information Science \& Technology, University of Southern
  California, Los Angeles, California 90089, USA}
\affiliation{Information Sciences Institute, University of Southern California, Marina del
  Rey, CA 90292}

\author{Daniel A. Lidar}
\affiliation{Department of Physics and Astronomy, University of Southern California, Los
  Angeles, California 90089, USA}
\affiliation{Center for Quantum  Information Science \& Technology, University of Southern
  California, Los Angeles, California 90089, USA}
\affiliation{Department of Electrical Engineering, University of Southern California, Los
  Angeles, California 90089, USA}
\affiliation{Department of Chemistry, University of Southern California, Los Angeles,
  California 90089, USA}

\begin{abstract}
  Closed-system quantum annealing is expected to sometimes fail spectacularly in solving
  simple problems for which the gap becomes exponentially small in the problem size. Much
  less is known about whether this gap scaling also impedes open-system quantum
  annealing. Here we study the performance of a quantum annealing processor in solving
  such a problem: a ferromagnetic chain with sectors of alternating coupling strength that
  is classically trivial but exhibits an exponentially decreasing gap in the sector
  size. The gap is several orders of magnitude smaller than the device
  temperature. Contrary to the closed-system expectation, the success probability rises
  for sufficiently large sector sizes. The success probability is strongly correlated with
  the number of thermally accessible excited states at the critical point. We demonstrate
  that this behavior is consistent with a quantum open-system description that is
  unrelated to thermal relaxation, and is instead dominated by the system's properties at
  the critical point.
\end{abstract}

\maketitle
\section{Introduction}
\noindent Quantum annealing (QA)~\cite{apolloni_1989_quantumstochastic,
  apolloni_1990_numericalimplementation,ray_1989_skmodel,somorjai_1991_novelapproach,
  amara_1993_globalenergy,finnila_1994_quantumannealing,kadowaki_1998_quantumannealing,
  das_2008_colloquiumquantum}, also known as the quantum adiabatic
algorithm~\cite{farhi_2000_quantumcomputation,farhi_2001_quantumadiabatic} or adiabatic
quantum optimization~\cite{smelyanskiy_2001_simulationsadiabatic,
  reichardt_2004_quantumadiabatic} is a heuristic quantum algorithm for solving
combinatorial optimization problems. Starting from the ground state of the initial
Hamiltonian, typically a transverse field, the algorithm relies on continuously deforming
the Hamiltonian such that the system reaches the final ground state--typically of a
longitudinal Ising model---thus solving the optimization problem. In the closed-system
setting, the adiabatic theorem of quantum mechanics~\cite{kato_1950_adiabatictheorem}
provides a guarantee that QA will find the final ground state if the run-time is
sufficiently large relative to the inverse of the quantum ground state energy
gap~\cite{jansen_2007_boundsadiabatic,lidar_2009_adiabaticapproximation}.  However, this
does not guarantee that QA will generally perform better than classical optimization
algorithms. In fact, it is well known that QA, implemented as a transverse field Ising
model, can result in dramatic slowdowns relative to classical algorithms even for very
simple optimization problems \cite{dam_2001_howpowerful,reichardt_2004_quantumadiabatic,
  farhi_2008_howmake,jorg_2008_simpleglass,laumann_2012_quantumadiabatic}. Generally, this
is attributed to the appearance of exponentially small gaps in such
problems~\cite{albash_2018_AQC}.

A case in point is the ferromagnetic Ising spin chain with alternating coupling strength
and open boundary conditions studied by Reichardt~\cite{reichardt_2004_quantumadiabatic}. The
`alternating sectors chain' (ASC) of length $N$ spins is divided into equally sized
sectors of size $n$ of `heavy' couplings $W_{1}$ and `light' couplings $W_{2}$, with
$W_1 > W_2 > 0$. Since all the couplings are ferromagnetic, the problem is trivial to
solve by inspection: the two degenerate ground states are the fully-aligned states, with
all spins pointing either up or down. However, this simple problem poses a challenge for
closed-system QA since the transverse field Ising model exhibits an exponentially small
gap in the sector size $n$~\cite{reichardt_2004_quantumadiabatic}, thus forcing the run-time to be
exponentially long in order to guarantee a constant success probability. A related result
is that QA performs exponentially worse than its imaginary-time counterpart for disordered
transverse field Ising chains with open boundary
conditions~\cite{zanca_2016_quantumannealing}, where QA exhibits an infinite-randomness
critical point~\cite{fisher_1995_criticalbehavior}.

As a corollary, we may naively expect that for a fixed run-time, the success probability
will decrease exponentially and monotonically with the sector size. While such a
conclusion does not follow logically from the adiabatic theorem, it is supported by the
well-studied Landau-Zener two-level problem~\cite{landau_1932_zurtheorie,zener_1932_nonadiabaticcrossing,joye_1994_prooflandauzener}.
How relevant are such dire closed-system expectations for real-world devices? By varying
the sector size of the ASC problem on a physical quantum annealer, we find a drastic
departure from the above expectations. Instead of a monotonically decreasing success
probability (at constant run-time), we observe that the success probability starts to grow
above a critical sector size $n^\ast$, which depends mildly on the chain parameters
$(W_1,W_2)$. We explain this behavior in terms of a simple open-system model whose salient
feature is the number of thermally accessible states from the instantaneous ground state
at the quantum critical point. The scaling of this `thermal density of states' is
non-monotonic with the sector size and peaks at $n^\ast$, thus strongly correlating with the
success probability of the quantum annealer. Our model then explains the success
probability behavior as arising predominantly from the number of thermally accessible
excitations from the ground state, and we support this model by adiabatic master equation
simulations.

Our result does not imply that open-system effects can lend an advantage to QA, and hence
it is different from proposed mechanisms for how open system effects can assist QA. For
example, thermal relaxation is known to provide one form of assistance to
QA~\cite{childs_2001_robustnessadiabatic,sarandy_2005_adiabaticquantum,amin_2008_thermallyassisted,venuti_2017_relaxationadiabatic},
but our model does not use thermal relaxation to increase the success probability above
$n^\ast$. We note that Ref.~\cite{amin_2008_thermallyassisted} introduced the idea that significant mixing due to
open system effects (beyond relaxation) at an anti-crossing between the first excited and
ground states could provide an advantage, and its theoretical predictions were supported
by the experiments in Ref.~\cite{dickson_2013_thermallyassisted}. In Ref.~\cite{amin_2008_thermallyassisted} an analysis of adiabatic
Grover search was performed (a model which cannot be experimentally implemented in a
transverse field Ising model), along with numerical simulations of random field Ising
models. In contrast, here we treat an analytically solvable model that is also
experimentally implementable using current quantum annealing hardware.

We also compare our empirical results to the predictions of the classical spin-vector
Monte Carlo model~\cite{shin_2014_howquantum}, and find that it does not adequately explain
them. Our study lends credence to the notion that the performance of real-world QA devices
can differ substantially from the scaling of the quantum gap.

\section{Result}
\subsection{The alternating sectors chain model}
\begin{figure}
  \centering
  \includegraphics[width=0.5\textwidth]{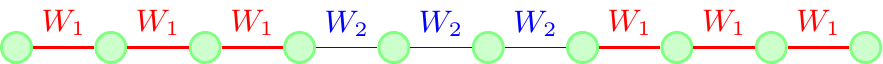}
  \caption{\textbf{Illustration of an alternating sector chain (ASC)}. This example has
      sector size $n=3$, length $N=10$ and number of sectors $2b+1=3$. Red lines denote the
      heavy sector with coupling $W_{1}$, blue lines denote the light sector with coupling
      $W_{2}<W_1$.}
\label{fig:1}
\end{figure}

\noindent We consider the transverse Ising model with a time-dependent
Hamiltonian of the form:
\begin{equation}
  \label{eq:1}
  H(s) = -A(s) \sum_{i=1}^N \sigma_{i}^{x} + B(s) H_{\text{ASC}}\ ,
\end{equation}
where $t_{\mathrm{f}}$ is the total annealing time, $s=t/t_{\mathrm{f}}\in [0,1]$, and
$A(s)$ and $B(s)$ are the annealing schedules, monotonically decreasing and increasing,
respectively, satisfying $B(0)=0$ and $A(1)=0$. The alternating sectors chain
Hamiltonian is
\begin{equation}
  \label{eq:2}
  H_{\text{ASC}} = -\sum_{i=1}^{N-1} J_{i} \sigma_{i}^{z}\sigma_{i+1}^{z}\ ,
\end{equation}
where for a given sector size $n$ the couplings are given by
\begin{equation}
  \label{eq:3}
  J_{i} =
  \begin{cases}
    W_{1} &\quad \text{if } \lceil i/n\rceil \text{ is odd} \\
    W_{2} &\quad \text{otherwise}
  \end{cases}
\end{equation}
Thus the $b+1$ odd-numbered sectors are `heavy' ($J_i=W_{1}$), and the
$b$ even-numbered sectors are `light' ($J_i=W_{2}$) for a total of $2b+1=\frac{N-1}{n}$
sectors. This is illustrated in Fig.~\ref{fig:1}.

\begin{figure*}
  \includegraphics[width=\textwidth]{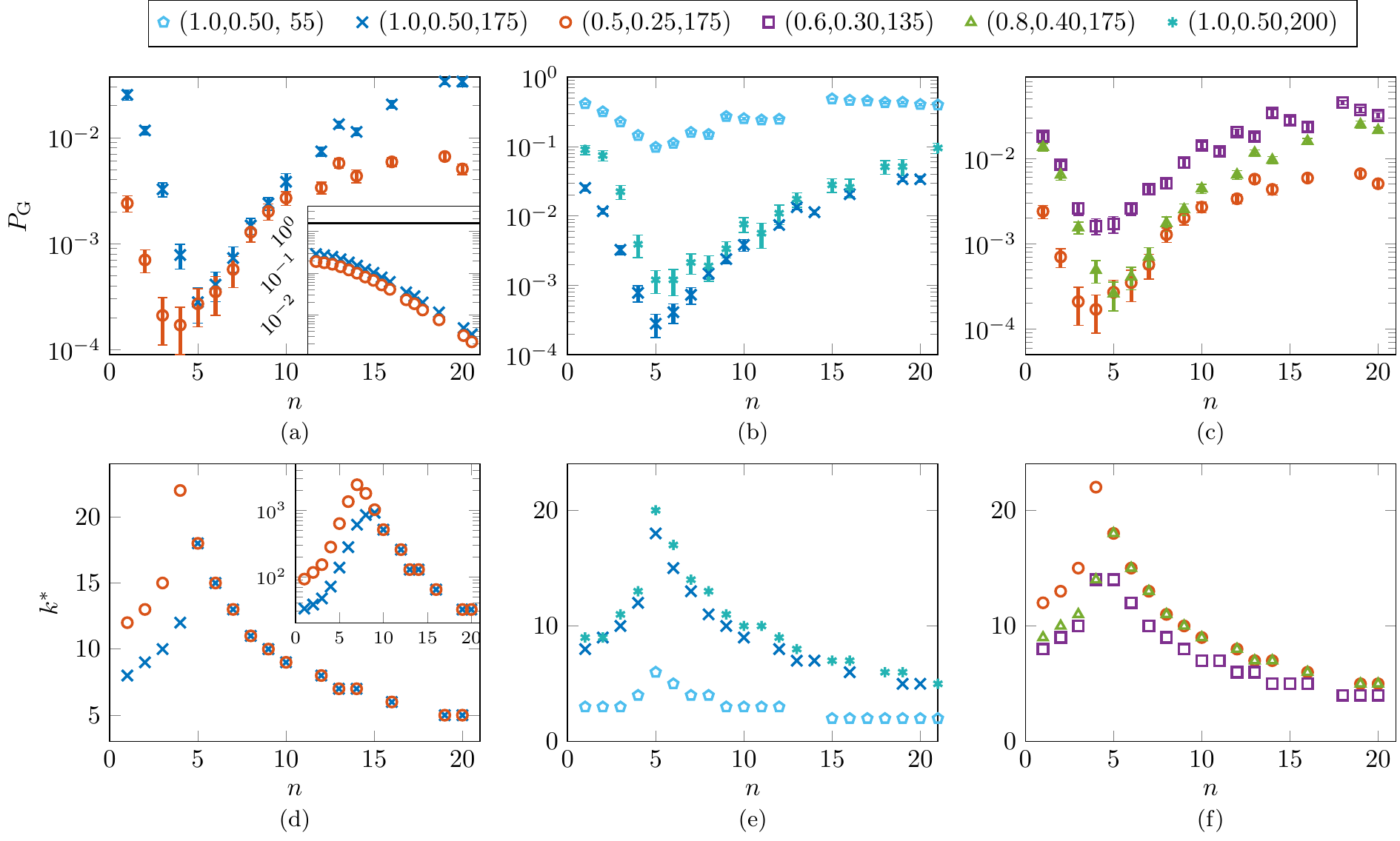}
  \caption{\textbf{Empirical success probability vs $\boldsymbol{k}^{\ast}$ for the ASC
      problem on the DW2X processor.} $k^{\ast}$ denotes the number of single-fermion
    energies that fall below the thermal energy gap at the point of the minimum gap
    $s^\ast$. The legend entries indicate the chain parameters: ($W_{1}$,$W_{2}$,$N$). The
    error bars everywhere indicate $95\%$ confidence intervals calculated using a
    bootstrap over different gauges and embeddings. {\bf (a)-(c)} Contrary to
    closed-system theory expectations, the success probability $P_\text{G}$ is
    non-monotonic in the sector size $n$, first decreasing and then increasing,
    exponentially. \textbf{Inset (a)}: The minimum gap (in GHz) of the chains as a
    function of the sector size $n\in\{1,\dots,20\}$. The solid black line denotes the
    operating temperature energy scale of the DW2X. {\bf (d)-(f)} For all chains we
    studied the ground state success probability has a minimum at the sector size $n^*$
    where the peak in the number of single-fermion states $k^{\ast}$ occurs [compare with
    (a)-(c)]. The rise and fall pattern, as well as the location of $n^*$, are in
    agreement with the behavior of $P_\text{G}$ within the error bars. {\bf Inset (d)}:
    The total number of energy eigenstates that fall below the thermal energy gap as a
    function of the sector size $n$. In this case the peak position does not agree with
    the ground state success probability minimum.}
  \label{fig:2}
\end{figure*}
We briefly summarize the intuitive argument of Ref.~\cite{reichardt_2004_quantumadiabatic} for the
failure of QA to efficiently solve the ASC problem. Consider the $N\gg 1$ and $n\gg 1$ limit,
where any given light or heavy sector resembles a uniform transverse field Ising
chain. Each such transverse field Ising chain encounters a quantum phase transition
separating the disordered phase and the ordered phase when the strength of the transverse
field and the chain coupling are equal, i.e., when
$A(s)=B(s)J_i$~\cite{sachdev_2011_quantumphase}. Therefore the heavy sectors order
independently before the light sectors during the anneal. Since the transverse field
generates only local spin flips, QA is likely to get stuck in a local minimum with domain
walls (antiparallel spins resulting in unsatisfied couplings) in the disordered (light)
sectors, if $t_{\mathrm{f}}$ is less than exponential in $n$. We note that this mechanism,
in which large local regions order before the whole is well-known in disordered,
geometrically local optimization problems, giving rise to a Griffiths
phase~\cite{fisher_1995_criticalbehavior}.

This argument explains the behavior of a closed-system quantum annealer operating in the
adiabatic limit. To check its experimental relevance, we next present the results of tests
performed with a physical quantum annealer operating at non-zero temperature.

\subsection{Empirical results}
\noindent As an instantiation of a physical quantum annealer we used a D-Wave 2X (DW2X)
processor. We consider ASCs with sector size $n\in[2,20]$. Since the number of sectors
$b = (N-1)/n$ must be an integer, the chain length varies slightly with $n$. The minimum
gap for these chains is below the processor temperature. Additional details about the
processor and of our implementation of these chains are given in Methods.

Figures~\ref{fig:2}(a)-\ref{fig:2}(c) show the empirical success probability results for a
fixed annealing time $t_{\mathrm{f}}=5\mu$s.
Longer annealing times do not change the qualitative behavior of the results, but do lead
to changes in the success probability (we provide these results
in~\cref{note:7}). A longer annealing time can result in more thermal
excitations near the minimum gap, but it may also allow more time for ground state
repopulation after the minimum gap. The latter can be characterized in terms of a
recombination of fermionic excitations by a quantum-diffusion mediated process
\cite{smelyanskiy_2017_quantumannealing}. Unfortunately, we cannot distinguish between these two
effects, as we only have access to their combined effect in the final-time success
probability.

In stark contrast to the theoretical closed-system expectation, the success probability
does not decrease monotonically with sector size, but exhibits a minimum, after which it
grows back to close to its initial value. The decline as well as the initial rise are
exponential in $n$. Longer chains result in a lower $P_\text{G}$ and a more pronounced
minimum, but the position of the minimum depends only weakly on the chain parameter values
$(W_1, W_2)$ (the value of $n^*$ shifts to the right as $(W_1,W_2)$ are increased) but not
on $N$.

What might explain this behavior? Clearly, a purely gap-based approach cannot suffice,
since the gap shrinks exponentially in $n$ for the ASC
problem~\cite{reichardt_2004_quantumadiabatic} [see also the inset of
Fig.~\ref{fig:2}(a)]. However, for all chain parameters we have studied, the temperature
is greater than the quantum minimum gap. In this setting not only the gap matters, but
also the number of accessible energy levels that fall within the energy scale set by the
temperature. In an open-system description of quantum annealing~\cite{childs_2001_robustnessadiabatic,amin_2009_decoherenceadiabatic,albash_2012_quantumadiabatic,deng_2013_decoherenceinduced,ashhab_2014_landauzenertransitions,albash_2015_decoherenceadiabatic},
both the Boltzmann factor $\exp(-\beta \Delta)$ ($\beta$ denotes the inverse temperature and
$\Delta$ is the minimum gap) and the density of states determine the excitation and relaxation
rates out of and back to the ground state. As we demonstrate next, the features of the
DW2X success probability results, specifically the exponential fall and rise with $n$, and
the position of the minimum, can be explained in terms of the number of single-fermion
states that lie within the temperature energy scale at the critical point.

\subsection{Fermionization}
\noindent We can determine the spectrum of the quantum Hamiltonian [Eq.~\eqref{eq:1}] by
transforming the system into a system of free fermions with fermionic raising and lowering
operators $\eta_k^\dagger$ and $\eta_k$~\cite{lieb_1961_twosoluble,sachdev_2011_quantumphase}. The result
is~\cite{reichardt_2004_quantumadiabatic}:
\begin{equation}
  \label{eq:4}
  H(s) = E_{\mathrm{g}}(s) + \sum_{k=1}^{N} \lambda_{k}(s) \eta_{k}^{\dagger}\eta_{k} \ ,
\end{equation}
where $E_{\mathrm{g}}(s)$ is the instantaneous ground state energy and $\{\lambda_{k}(s)\}$ are the
single-fermion state energies, i.e., the eigenvalues of the linear system
\begin{equation}
  \label{eq:5}
\vec{\Phi}_{k}(s)(\mytensor{A}-\mytensor{B})(\mytensor{A}+\mytensor{B}) = \lambda_{k}^{2}(s)\vec{\Phi}_{k}(s)\ ,
\end{equation}
where the matrices $\mytensor{A}$ and $\mytensor{B}$ are tridiagonal and are given
in~\cref{note:1} along with full details of the derivation. The vacuum of the
fermionic system $\ket{0}$ is defined by $\eta_{k}\ket{0}=0\ \forall\ k$ and is the ground state of
the system. Higher energy states correspond to single and many-particle fermionic
excitations of the vacuum. At the end of the anneal, fermionic excitations corresponds to
domain walls in the classical Ising chain (see~\cref{note:2}).

The Ising problem is $\mathbb{Z}_{2}$-symmetric, so the ground state and the first excited
state of the quantum Hamiltonian merge towards the end of evolution to form a doubly
degenerate ground state. Since any population in the instantaneous first excited state
will merge back with the ground state at the end of the evolution, the relevant minimum
gap of the problem is the gap between the ground state and the second excited state:
$\Delta(s) = \lambda_{2}(s)$, which occurs at the point $s^*=\argmin_{s\in[0,1]}\Delta(s)$.
In the thermodynamic limit, this point coincides with the quantum critical point where the
geometric mean of the Ising fields balances the transverse field,
$A(s^{\ast})=\sqrt{W_{1}W_{2}}B(s^{\ast})$~\cite{hermisson_1997_aperiodicising,pfeuty_1979_exactresult}.
Henceforth we write $\Delta \equiv \Delta(s^*)$ for the minimum gap.

\subsection{Spectral analysis}
\noindent Let $k^{\ast}$ be the number of single-fermion states with energy smaller than the
thermal gap at the critical point, i.e.,
\begin{equation}
  \label{eq:6}
  k^{\ast} = \argmax_k \Big\{ \lambda_{k}(s^{\ast}) < T \Big\}\ .
\end{equation}
As can be seen by comparing Figs.~\ref{fig:2}(d)-\ref{fig:2}(f) to
Figs.~\ref{fig:2}(a)-\ref{fig:2}(c), we find that the behavior of $k^{\ast}$ correlates
strongly with the ground state success probability for all ASC cases we tested, when we
set $T=12\ \text{mK}=1.57$\ GHz, the operating temperature of the DW2X processor (we use
$k_{\mathrm{B}} = \hbar = 1$ units throughout). Specifically, $k^\ast$ peaks exactly where the success
probability is minimized, which strongly suggests that $k^{\ast}$ is the relevant quantity
explaining the empirically observed quantum annealing success probability. Longer chains
result in a larger value of $k^{\ast}$ and a more pronounced maximum. Of all the ASC sets we
tried, we only found a partial exception to this rule for the case $(1,0.5,200)$, where
$k^{\ast}$ peaks at $n^{\ast}=5$ [Fig.~\ref{fig:2}(e)] but the empirical success probability for
$n=5$ and $n=6$ is roughly the same [Fig.~\ref{fig:2}(b)]. We show later that this
exception can be resolved when the details of the energy spectrum are taken into account
via numerical simulations.

In contrast, the total number of energy eigenstates (including multi-fermion states) that
lie within the thermal gap $[E_{\textrm{g}}(s^{\ast}) , E_{\textrm{g}}(s^\ast)+T]$, while rising
and falling exponentially in $n$ like the empirical success probability in
Fig.~\ref{fig:2}(a), does not peak in agreement with the peak position of the latter [see
the inset of Fig.~\ref{fig:2}(d)].\\
\begin{figure}
  \centering
  \includegraphics[width=\columnwidth,height=0.75\columnwidth]{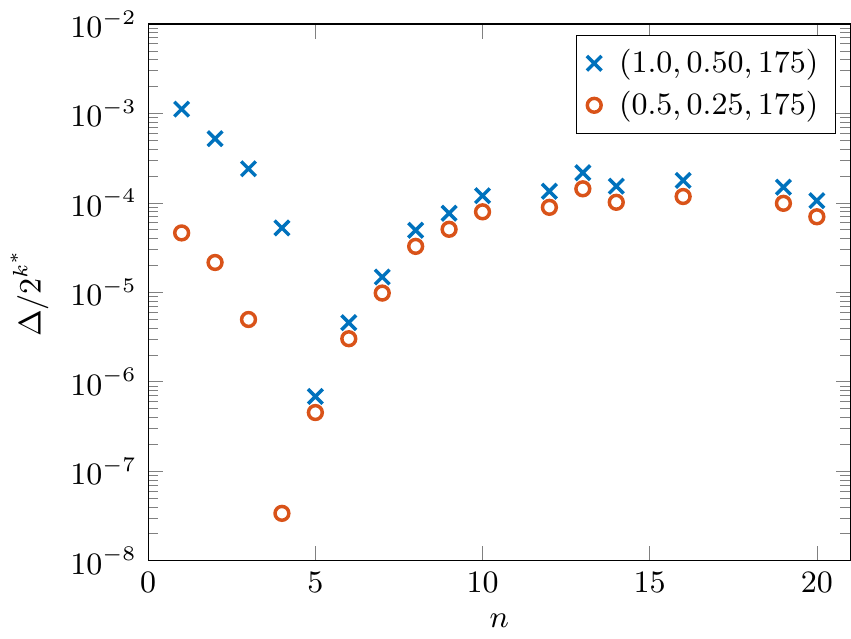}
  \caption{\textbf{Ratio of the gap to the thermal density of states, as a function of
      sector size}. Two alternating sector chain cases are shown. The position of the
    minimum is determined by $d$ rather than $\Delta$, as can be seen by comparing to
    Fig.~\ref{fig:2}(d) where the plot of $d=2^{k^{\ast}}$ alone correlates well with the
    position of minima in the empirical success probability curves.}
  \label{fig:3}
\end{figure}

Why and how does the behavior of $k^*$ explain the value of $n^*$? Heuristically, we
expect the success probability to behave as
\begin{equation}
  \label{eq:7}
  P_{\mathrm{G}} \sim \frac{1-e^{-\beta \Delta}}{d}\ ,
\end{equation}
where $d$ is the `thermal density of states' at the critical point $s^*$. Note that the
role of the gap here is different from the closed-system case, since we are assuming that
thermal transitions dominate over diabatic ones, so that the gap is compared to the
temperature rather than the annealing time. Contrast this with the closed system case,
where the Landau-Zener formula for closed two-level systems and Hamiltonians analytic in
the time parameter (subject to a variety of additional technical conditions) states that:
$P_{\mathrm{G}} \sim 1-e^{-\eta\Delta^2 t_{\mathrm{f}}}$, where $\eta$ is a constant with units of time
that depends on the parameters that quantify the behavior at the avoided crossing
(appearing in, e.g., the proof of Theorem 2.1 in
Ref.~\cite{joye_1994_prooflandauzener}). Since then
$P_{\mathrm{G}} = O(\eta\Delta^2 t_{\mathrm{f}})$, we expect the success probability to decrease
exponentially at constant run-time $t_{\mathrm{f}}$ if the gap shrinks exponentially in
the system size.

Our key assumption is that the thermal transitions between states differing by more than
one fermion are negligible. That is, thermal excitation (relaxation) only happens via
creation (annihilation) of one fermion at a time (see~\cref{note:3} for a detailed
argument). Additionally, the Boltzmann factor suppresses excitations that require energy
exchange greater than $\lambda_{k^{\ast}}$. Starting from the ground state, all single-fermion
states with energy $\leq E_{\mathrm{g}} + \lambda_{k^{\ast}}$ are populated first, followed by all
two-fermion states with total energy
$\leq E_{\mathrm{g}}+ \lambda_{k^{\ast}}+\lambda_{k^{\ast}-1}$, etc. In all,
$\sum_{k=1}^{k^*} \binom{k^*}{k} = 2^{k^{\ast}}-1$ excited states are thermally populated in
this manner. Thus $d~\sim 2^{k^{\ast}}$ states are thermally accessible from the ground state.

For a sufficiently small gap we have $1-e^{-\beta\Delta}\sim \beta\Delta$, so that
$P_{\mathrm{G}}\sim \beta\Delta/d$. As can be seen from Figs.~\ref{fig:2}(d)-\ref{fig:2}(f),
$k^*$ rises and falls steeply for $n<n^*$ and $n>n^*$ respectively. For the ASCs under
consideration, $d$ varies much faster with $n$ than the gap $\Delta$ (see
Fig.~\ref{fig:3}). Thus $P_{\mathrm{G}}\sim 2^{-k^{\ast}}$. This argument explains both the
observed minimum of $P_{\mathrm{G}}$ at $n^*$ and the exponential drop and rise of
$P_{\mathrm{G}}$ with $n$, in terms of the thermal density of states. In~\cref{note:4}
we give a more detailed argument based on transition rates obtained
from the adiabatic master equation, which we discuss next.
\subsection{Master equation model}
\noindent We now consider a simplified model of the open system dynamics in order to make
numerical predictions. We take the evolution of the populations $\vec{p}=\{p_a\}$ in the
instantaneous energy eigenbasis of the system to be described by a Pauli master
equation~\cite{pauli_1928_htheorem}. The form of the Pauli master equation is identical to
that of the adiabatic Markovian quantum master equation~\cite{albash_2012_quantumadiabatic},
derived for a system of qubits weakly coupled to independent identical bosonic baths. The
master equation with an Ohmic bosonic bath has been successfully applied to qualitatively
(and sometimes quantitatively) reproduce empirical D-Wave
data~\cite{albash_2015_consistencytests,boixo_2013_experimentalsignature,albash_2015_reexaminationevidence,albash_2015_reexaminingclassical}.
However, it does not account for $1/f$ noise~\cite{yoshihara_2006_decoherenceflux}, which may
invalidate the weak coupling approximation when the energy gap is smaller than the
temperature~\cite{boixo_2016_computationalmultiqubit}.
\begin{figure}
  \subfigure[\ ]{\includegraphics[width=\columnwidth,height=0.618\columnwidth]
    {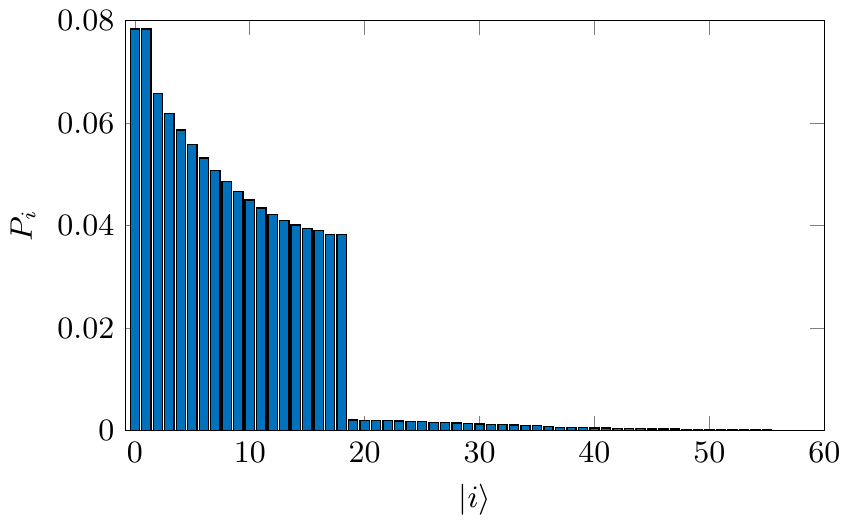}\label{fig:4a}}
  \subfigure[\ ]{\includegraphics[width=\columnwidth,height=0.618\columnwidth]
    {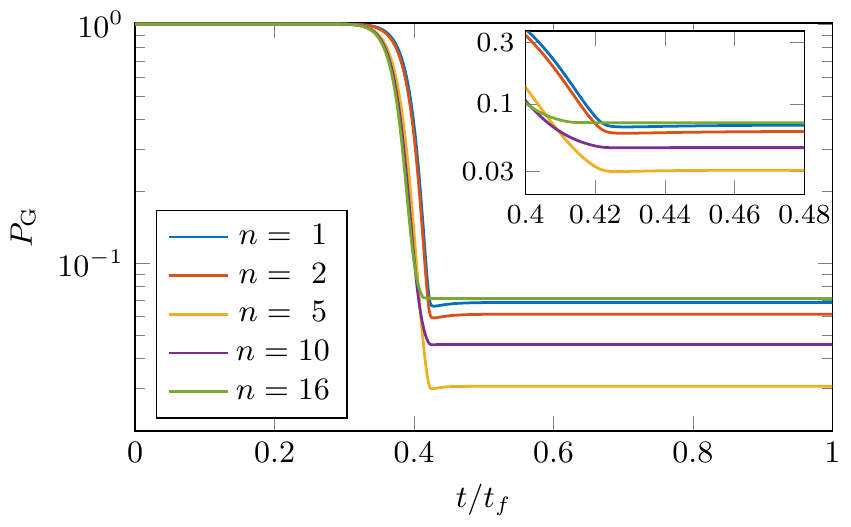}\label{fig:4b}}
  \caption{\textbf{Master equation results for the state populations when restricting the
      excited states to single-fermion states.} (a) The population in each single-fermion
    state at $t=t_{\mathrm{f}}$ in a one-fermion simulation. The chain parameters are
    $N=176$, $W_1 = 1$, $W_2=0.5$, $t_{\mathrm{f}}=5 \mu\text{s}$, and $n=5$. With the
    annealing schedule given in Methods, the quantum minimum gap is at
    $s^{\ast}=t^{\ast}/t_{\mathrm{f}}\approx0.424$. At this point we find $k^{\ast}=18$ single-fermion
    states below the thermal energy $T=12\text{mK}$ (D-Wave processor operating
    temperature). As expected, in one-fermion simulations, most of the population is found
    in the first $k^*$ states. A long tail of more energetic single particle states beyond
    the first $k^*$ retain some population. (b) Evolution of the instantaneous ground
    state populations for ASCs with the same parameters as in (a), but for different
    sector sizes $n$ and with two-fermion states. The ground state loses the majority of
    its population as it approaches the minimum gap point at $t/t_{\mathrm{f}}=s^{\ast}$. The
    largest drop is found for $n=n^*=5$. Inset: Magnification of the region around the
    minimum gap. Relaxation plays essentially no role. Instead, the population freezes
    almost immediately.}
  \label{fig:4}
\end{figure}

After taking diagonal matrix elements and restricting just to the dissipative
(non-Hermitian) part one obtains the Pauli master equation~\cite{pauli_1928_htheorem}
describing the evolution of the population $\vec{p}=\{p_a\}$ in the instantaneous energy
eigenbasis of the system~\cite{albash_2015_decoherenceadiabatic}:
\begin{equation}
  \label{eq:8}
  \frac{\partial {p}_{a} }{\partial t}= \sum_{b \neq a} \gamma(\omega_{ba}) M_{ab} p_{b} -  \sum_{b \neq a} \gamma(\omega_{ab}) M_{ba}
  p_{a}\ .
\end{equation}
Here all quantities are time-dependent and the matrix elements are
\begin{equation}
  \label{eq:9}
  M_{ab}(s) = \sum_{\alpha=1}^N |\braket{a(s)|\sigz_{\alpha}|b(s)}|^{2}\ ,
\end{equation}
where we have assumed an independent thermal bath for each qubit $\alpha$ and where the indices
$a$ and $b$ run over the instantaneous energy eigenstates of the system
Hamiltonian (\cref{eq:4}) in the fermionic representation [i.e.,
$H(s)\ket{a(s)} = E_a(s)\ket{a(s)}$] and $\omega_{ab}=E_{a}-E_{b}$ is the corresponding
instantaneous Bohr frequency. Since the basis we have written this equation in is
time-dependent, there are additional terms associated with the changing basis
\cite{albash_2012_quantumadiabatic}, but we ignore these terms here since we are assuming that the
system is dominated by the dissipative dynamics associated with its interaction with its
thermal environment.
\begin{figure*}
  \includegraphics[width=\textwidth]{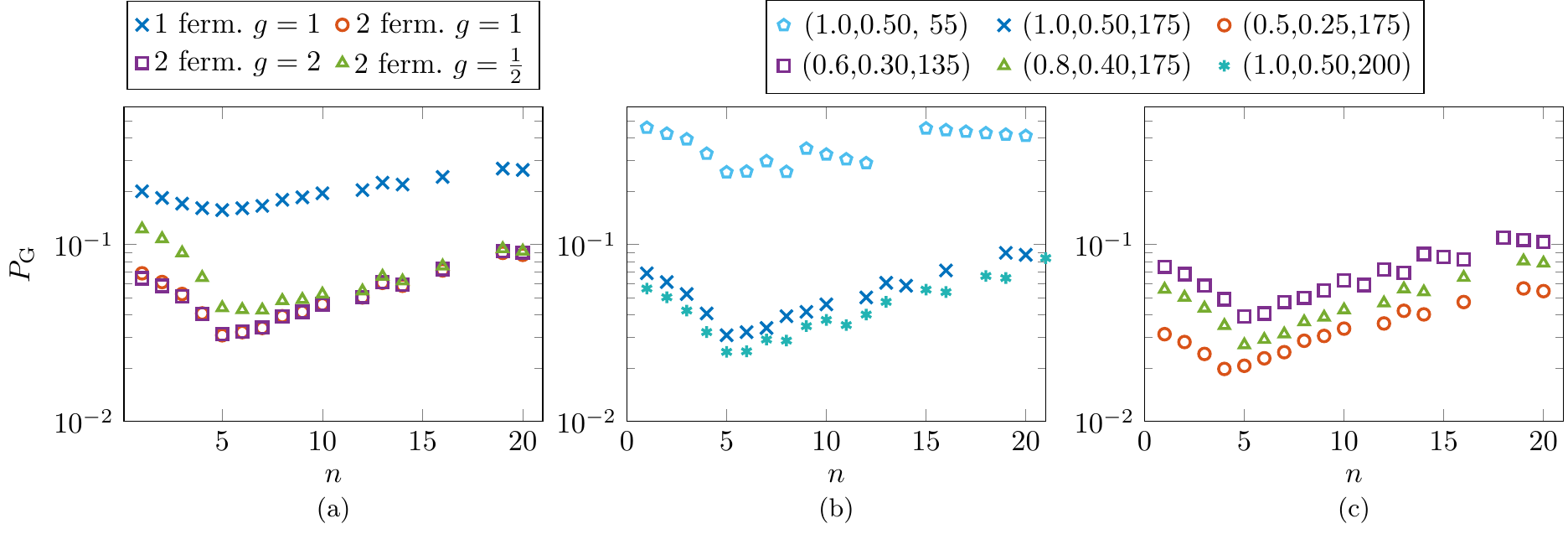}
  \caption{\textbf{Master equation results for the ground state population when
      restricting the excited states to single and two fermion states.} (a) The result of
    simulating the ASC problem with parameters $(1,0.5,175)$ via the adiabatic Pauli
    master equation~\eqref{eq:8}, restricted to the vacuum + single-fermion states, and
    vacuum + single-fermion + two-fermion states. Also shown is the dependence on the
    system-bath coupling parameter $g$ in the two-fermion case; doubling it has little
    impact, whereas halving it increases the success probability somewhat for $n<14$. The
    position of the minimum at $n^*=5$ matches the empirical result seen in
    Fig.~\ref{fig:2}(a), except when $g=1/2$, i.e., the position is robust to doubling $g$
    but not to halving it. Panels (b) and (c) show additional $2$-fermion master equation
    results with $g=1$. Note that for the $(1,0.5,200)$ chain, these simulations exhibit
    better agreement with the DW2X data than the simple $k^*$ analysis plotted in
    Figs.~\ref{fig:2}(d)-\ref{fig:2}(f). This is because the simulations also keep track
    of the Boltzmann factor.}
  \label{fig:5}
\end{figure*}

The rates $\gamma(\omega)$ satisfy the quantum detailed balance condition~\cite{haag_1967_equilibriumstates,breuer_2002_theoryopen},
$\gamma(-\omega) = e^{-\beta \omega} \gamma(\omega)$, where $\omega \geq 0$. In our model each qubit is coupled to an
independent pure-dephasing bath with an Ohmic power spectrum:
\begin{equation}
\label{eq:10}
\gamma(\omega) = \ 2\pi\eta g^{2} \frac{\omega e^{-|\omega|/\omega_{c}}}{1-e^{-\beta \omega}}\ ,
\end{equation}
with UV cutoff $\omega_{c}=8\pi$ GHz and the dimensionless coupling constant
$\eta g^{2}=1.2\times 10^{-4}$. The choice for the UV cutoff satisfies the assumptions made in the
derivation of the master equation in the Lindblad form~\cite{albash_2012_quantumadiabatic}. Note
that we do not adjust any of the master equation parameter values, which are taken from
Ref.~\cite{albash_2015_consistencytests}. Details about the numerical solution procedure are
given in Methods, and in~\cref{note:5} we also confirm that the validity conditions for the
derivation of the master equation are satisfied for a relevant range of $n$ values given
the parameters of our empirical tests.

Numerically solving the master equation while accounting for all thermally populated
$2^{k^{\ast}}$ states is computationally prohibitive, but we can partly verify our
interpretation by restricting the evolution of the system described in Eq.~\eqref{eq:8} to
the vacuum and single-fermion states. This is justified in~\cref{note:3},
where we show that transitions between states differing by more than a single fermion are
negligible. In other words, the dominant thermal transitions occur from the vacuum to the
single-fermion states, from the single-fermion states to the two-fermion states, etc. The
restriction to the vacuum and single-fermion states further simplifies the master
equation~\eqref{eq:8} to:
\begin{align}
  \dot{p}_{0} &= \sum_{b} \gamma(\lambda_{b}) M_{b} p_{b} -   p_{0}\sum_{b} \gamma(-\lambda_{b}) M_{b} \label{eq:11}  \\
  \dot{p}_{i} &= \gamma(-\lambda_{i})M_{i}p_{0} - \gamma(\lambda_{i})M_{i}p_{i}\ , \label{eq:12}
\end{align}
where $\left\{p_{b} \right\}_{b=1}^N$ are the single-particle fermion energy populations
and $\left\{\lambda_{b} \right\}$ their energies found by solving Eq.~\eqref{eq:5}, and
$M_{ab}$ [Eq.~\eqref{eq:9}] becomes
$M_{b} = \sum_{\alpha=1}^N |\braket{0|\sigz_{\alpha}|b}|^{2}$. For a better approximation that accounts
for more states, we can also perform a two-fermion calculation where we keep the vacuum,
the first $k^{\ast}$ one fermion-states and the next $k^{\ast}(k^{\ast}-1)/2$ two fermion
states. For two-fermion simulations the master equation becomes~\cref{eq:11,eq:12} along
with

\begin{align}
  \dot{p}_{i} ={}& \gamma(-\lambda_{i})M_{i}p_{0} - \gamma(\lambda_{i})M_{i}p_{i} \notag  \\
                 & + \sum_{j\neq i}\gamma(\lambda_{j})M_{j}p_{ij} - p_{i}\sum_{j\neq
                   i}\gamma(-\lambda_{i})M_{j} \label{eq:13}\\
  \dot{p}_{ij} ={}& \gamma(-\lambda_{i})M_{i}p_{j} +\gamma(-\lambda_{j})M_{j}p_{i} - \gamma(\lambda_{i})M_{i}p_{ij} \notag \\
                 &- \gamma(\lambda_{j})M_{j}p_{ij}\ , \label{eq:14}
\end{align}
where all summations run from $1$ to $k^{\ast}$, and $p_{ij}$ denotes the population in the
two-particle fermion energy state $\eta_i^\dagger \eta_j^\dagger \ket{0}$.\\

We can now numerically solve this system of equations. As seen in Fig.~\ref{fig:4a},
where we plot the final populations in the different single particle fermion states at
$t=t_{\mathrm{f}}$ for one-fermion simulations, only the first $k^*$ single-fermion levels are
appreciably populated. This agrees with our aforementioned assumption that states with
energy greater than $\lambda_{k^*}$ are not thermally populated. In Fig.~\ref{fig:4b} we
plot the population in the instantaneous ground state as a function of time for
two-fermion simulations. The system starts in the gapped phase where the ground state
population is at its chosen initial value of $1$. The ground state rapidly loses
population via thermal excitation as the system approaches the critical point, after which
the population essentially freezes, with repopulation via relaxation from the excited
states essentially absent (see inset). Thus, it is not relaxation that explains the
increase in ground state population seen in Fig.~\ref{fig:2}(a)-(c) for $n>n^*$. Instead,
we find that the ground state population drops most deeply for $n=n^*$. This, in turn, is
explained by the behavior of $k^*$ seen in Fig.~\ref{fig:2}(d)-(f), as discussed earlier.

We show in Fig.~\ref{fig:5} the predicted final ground state population under the one and
two-fermion restriction. This minimal model already reproduces the correct location of the
minimum in $P_\text{G}$. It also reproduces the non-monotonic behavior of the success
probability. It does not correctly reproduce the exponential fall and rise. However,
including the two-fermion states gives the right trend: it leads to a faster decrease and
increase in the population without changing the position of the minimum, suggesting that a
simulation with the full $2^{k^{\ast}}$ states would recover the empirically observed
exponential dependence of the ground state population seen in
Fig.~\ref{fig:2}(a)-\ref{fig:2}(c).

\section{Discussion}
\noindent A commonly cited failure mode of closed-system quantum annealing is the
exponential closing of the quantum gap with increasing problem size. It is expected, on
the basis of the Landau-Zener formula and the quantum adiabatic theorem, that to keep the
success probability of the algorithm constant the run-time should increase
exponentially. As a consequence, one expects the success probability to degrade at
constant run-time if the gap decreases with increasing problem size. Our goal in this work
was to test this failure mode in an open-system setting where the temperature energy scale
is always larger than the minimum gap. We did so by studying the example of a
ferromagnetic Ising chain with alternating coupling-strength sectors, whose gap is
exponentially small in the sector size, on a quantum annealing device. Our tests showed
that while the success probability initially drops exponentially with the sector size, it
recovers for larger sector sizes. We found that this deviation from the expected
closed-system behavior is qualitatively and semi-quantitatively explained by the system's
spectrum around the quantum critical point. Specifically, the scaling of the quantum gap
alone does not account for the behavior of the system, and the scaling of the number of
energy eigenstates accessible via thermal excitations at the critical point (the thermal
density of states) explains the empirically observed ground state population.

Does there exist a classical explanation for our empirical results? We checked and found
that the spin vector Monte Carlo (SVMC) model~\cite{shin_2014_howquantum} is capable of matching
the empirical DW2X results provided we fine-tune its parameters for each specific chain
parameter set $\{W_1,W_2,N\}$. However, it does not provide as satisfactory a physical
explanation of the empirical results as the fermionic or master equation models, which
require no such fine-tuning; see~\cref{note:6} for details.

Our work demonstrates that care must be exercised when inferring the behavior of
open-system quantum annealing from a closed-system analysis of the scaling of the gap. It
has already been pointed out that quantum relaxation can play a beneficial
role~\cite{childs_2001_robustnessadiabatic,sarandy_2005_adiabaticquantum,amin_2008_thermallyassisted,dickson_2013_thermallyassisted,venuti_2017_relaxationadiabatic}.
However, we have shown that relaxation plays no role in the recovery of the ground state
population in our case. Instead, our work highlights the importance of a different
mechanism: the scaling of the number of thermally accessible excited states. Thus, to
fully assess the prospects of open-system quantum annealing, this mechanism must be
understood along with the scaling of the gap and the rate of thermal relaxation. Of
course, ultimately we only expect open-system quantum annealing to be scalable via the
introduction of error correction methods~\cite{young_2013_errorsuppression,
  jordan_2006_errorcorrecting,marvian_2017_subsystemcode,jiang_2017_noncommutingtwolocal}.

\section{Methods}
\begin{table}
  \centering
  \begin{tabular}{l|rgrgrgrgrgr}
    \hline
    $N$ & 174 & 175 & 172 & 173 & 176 & 175 & 176 & 169 \\
    \hline
    $n$ & 1   &  2  &  3  &  4  &  5  &  6  &  7  &  8  \\
    \hline\hline
    $N$ & 172 & 171 & 181 & 170 & 183 & 177 & 172 & 181 \\
    \hline
    $n$ & 9  & 10  & 12  & 13  & 14  & 16  & 19  & 20  \\
  \end{tabular}
  \caption{Chain length ($N$) and sector size ($n$) for $N\sim175$.
  \label{tab:1}}
\end{table}
\noindent\textbf{Alternating sector chains} We generated a set of ASCs with chains
lengths centered at $N\sim\{55,135,175,200\}$ and with sector sizes $n$ ranging from $2$ to
$20$. Since the chain length and sector size must obey the relation $(N-1)/n = 2b+1$ with
integer $b$, there is some variability in $N$. Table~\ref{tab:1}
gives the $(N,n)$ pair combinations we used for chain set with mean length $175$.\\

\begin{figure}
  \centering \includegraphics[width=0.4\textwidth,height=0.3\textwidth]{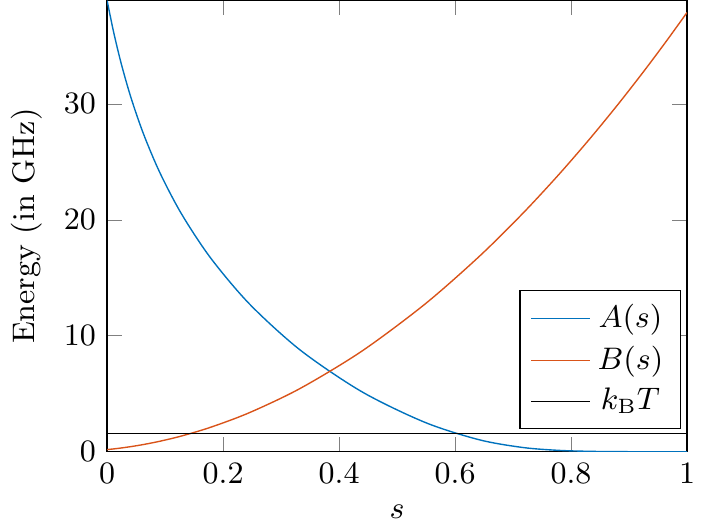}
  \caption{\textbf{Annealing schedules and temperature.} $A(s)$ and $B(s)$ are the
    annealing schedules of the D-Wave processor used in this work. The fridge temperature
    (horizontal black line) is $T=12$ mK.}
  \label{fig:6}
\end{figure}

\noindent\textbf{Quantum annealing processor used in this work} The D-Wave 2X processor
(DW2X) is an $1152$-qubit quantum annealing device made by D-Wave Systems, Inc., using
superconducting flux qubits~\cite{bunyk_2014_dwave}. The particular processor used in this
study is located at the University of Southern California's Information Sciences
Institute, with $1098$ functional qubits and an operating temperature of $12$ mK. The
total annealing time $t_{\mathrm{f}}$ can be set in the range $[5,2000]\ \mu$s. The time-dependent
Hamiltonian the processor is designed to implement is given by
\begin{equation}
  \label{eq:15}
  H(s) = A(s) \sum_{i} \sigma^{z}_{i} + B(s)\left( \sum_{i} h_{i}\sigma_{i}^{z} +
    \sum_{(i,j)} J_{ij} \sigma_{i}^{z}\sigma_{j}^{z} \right) \ ,
\end{equation}
with dimensionless time $s=t/t_{\mathrm{f}}$. Figure~\ref{fig:6} describes the annealing schedules
$A(s)$ and $B(s)$. The coupling strengths $J_{ij}$ between qubits $i$ and $j$ can be set
in the range $[-1,1]$ and the local fields $h_i$ can be set in the range $[-2,2]$.

We used $t_{\mathrm{f}}= 5\ \mu$s. For each ASC instance we implemented $10$ different embeddings,
with $10$ gauge transforms each~\cite{boixo_2014_evidencequantum}. In total, $10^5$ runs and
readouts were taken per instance. The reported success probability is defined as
the fraction of readouts corresponding to a correct ground state. For additional details
on the DW2X processor we used see, e.g., Ref.~\cite{albash_2018_demonstrationscaling}. \\

\noindent\textbf{Numerical procedure for solving the master equation} We solve the
coupled differential~\cref{eq:11,eq:12,eq:13,eq:14} using a fourth
order Runge-Kutta method given by Dormand-Prince~\cite{dormand_1980_familyembedded} with
non-negativity constraints~\cite{shampine_2005_nonnegativesolutions}. We compute the transition
matrix elements via~\cref{eq:50} and the bath correlation term via
Eq.~\eqref{eq:10}. \\

\noindent\textbf{Data Availability} The data that support the findings of this study are
available from the corresponding author upon reasonable request.

\acknowledgements

\noindent We thank Ben W.\ Reichardt for useful discussions. The computing resources used
for this work were provided by the USC Center for High Performance Computing and
Communications. A.M. was supported by the USC Provost Ph.D. Fellowship. The research is
based upon work (partially) supported by the Office of the Director of National
Intelligence (ODNI), Intelligence Advanced Research Projects Activity (IARPA), via the
U.S. Army Research Office contract W911NF-17-C-0050. The views and conclusions contained
herein are those of the authors and should not be interpreted as necessarily representing
the official policies or endorsements, either expressed or implied, of the ODNI, IARPA, or
the U.S. Government. The U.S. Government is authorized to reproduce and distribute
reprints for Governmental purposes notwithstanding any copyright annotation thereon.

\appendix

\section{Jordan-Wigner transform}
\label{note:1}

The Jordan-Wigner transform can be used to map a one-dimensional transverse field Ising
Hamiltonian to a Hamiltonian of uncoupled (free) fermions. A system of dimension
$2^{N}\times 2^{N}$ is thus effectively reduced to a system of dimension $N$, which is
essential for our simulations involving $N$ as large as $200$.  We briefly summarize the
approach found in the classic work of Lieb \textit{et al}.~\cite{lieb_1961_twosoluble}.

We start by defining fermionic raising operators
$a_{i}^{\dagger}=\frac{1}{2}\left(\otimes_{j=1}^{i-1}\sigx_{j}\right)(\sigz_{i}+\ii \sigy_{i})$
($i=1,\dots,N$) and their corresponding lowering operators $a_{i}$. It is easy to check
that the fermionic canonical commutation relations are then satisfied:
$\{a_{i},a_{j}\} = \{a_{i}^{\dagger},a_{j}^{\dagger}\}= 0$ and
$\{a^{\dagger}_{i},a_{j}\} = \delta_{ij}$. Rewriting~\cref{eq:1} as
\begin{equation}
  \label{eq:16}
  H = -\Gamma \sum_{i=1}^N \sigx_{i} - \sum_{i=1}^{N-1} J_{i}\sigz_{i}\sigz_{i+1}
\end{equation}
and substituting $\sigx_{i} = 1 - 2a^{\dagger}_{i}a_{i}$,
$\sigz_{i}\sigz_{i+1}=(a_{i}^{\dagger} - a_{i})(a_{i+1}^{\dagger} + a_{i+1})$ gives
\begin{equation}
 \label{eq:17}
  H = -N\Gamma + \sum_{ij} a_{i}^{\dagger}A_{ij}a_{j}
  + \frac{1}{2}(a_{i}^{\dagger}B_{ij}a_{j}^{\dagger} + a_{i}B_{ji}a_{j})\ ,
\end{equation}
or
\begin{equation}
  \label{eq:18}
  H = -N\Gamma + (\vec{a^\dagger})^T\mytensor{A}\vec{a} +
  \frac{1}{2}\left[(\mytensor{a^\dagger})^T\mytensor{B}\mytensor{a^\dagger} +
    \vec{a}^T\mytensor{B}^T\vec{a}\right]
\end{equation}
where $(\vec{a^\dagger})^T = (a_1^\dagger,\dots,a_{N}^\dagger)$, $\vec{a}^T = (a_1,\dots,a_{N})$, and
\begin{align}
    \mytensor{A} &=  \begin{pmatrix}
      2\Gamma     & -J_{1} &  0    &         &   \\
      -J_{1} & 2\Gamma    & -J_{2} &         &   \\
      0     & -J_{2} &  \ddots    & \ddots       & 0 \\
      &       & \ddots     & 2\Gamma       & -J_{N-1} \\
      & \cdots & & -J_{N-1} & 2\Gamma
    \end{pmatrix} \ , \label{eq:19} \\
    \mytensor{B} &= \begin{pmatrix}
      0     & -J_{1} &        &         &   \\
      J_{1} & 0      & -J_{2} &         &  \\
      & J_{2}  &  \ddots     & \ddots       & 0 \\
      &        & \ddots      & 0       & -J_{N-1} \\
      & \cdots & & J_{N-1} & 0
    \end{pmatrix}\ . \label{eq:20}
\end{align}

The Hamiltonian in~\cref{eq:17} is not fermion number conserving (it contains terms
such as $a_{i}^{\dagger}a_{j}^{\dagger}$, which means that  $\sum_{i} \sigx_{i}$ is not conserved), but
it can be diagonalized by a Bogoliubov transformation~\cite[Appendix A]{lieb_1961_twosoluble} in
terms of a new set of fermionic operators $\{ \eta_{i}, \eta_{i}^{\dagger}\}$:
\begin{equation}
  \label{eq:21}
  H = E_{\mathrm{g}} + \sum_{i=1}^{N} {\lambda_{i}}\eta_{i}^{\dagger}\eta_{i}\ .
\end{equation}
where the $\lambda_i$ are the single-fermion energies of the system. The new fermionic
operators are real linear combinations of the old ones:
\begin{align}
  \label{eq:22}
  \begin{split}
    \eta_{i}^{\dagger} &= \sum_k g_{ik}a^{\dagger}_{k} + \sum_k h_{ik} a_{k} \\
    \eta_{i} &= \sum_k g_{ik}a_{k} + \sum_k h_{ik} a^{\dagger}_{k}\ .
  \end{split}
\end{align}
From~\cref{eq:21} we get
\begin{equation}
  \label{eq:23}
  \left[ \eta_{k}, H \right] = \lambda_{k}\eta_{k}\ .
\end{equation}
Substituting~\cref{eq:17,eq:22} in~\cref{eq:23} and setting the
coefficients of every operator $\{a_{i},a^{\dagger}_{i}\}$ to zero gives
\begin{align}
  \label{eq:24}
  \begin{split}
    \lambda_{k}g_{ki} &= \sum_{j}g_{kj}A_{ji} - h_{kj}B_{ji} \\
    \lambda_{k}h_{ki} &= \sum_{j}g_{kj}B_{ji} - h_{kj}A_{ji}
  \end{split}
\end{align}
Let $\phi_{ik} = g_{ik} + h_{ik}$ and $\psi_{ik}=g_{ik}-h_{ik}$.  Plugging this into
~\cref{eq:24}, we get two coupled equations:
\begin{align}
  \vec{\Phi}_{k}(\mytensor{A}-\mytensor{B}) &= \lambda_{k}\vec{\Psi}_{k}  \label{eq:25} \\
  \vec{\Psi}_{k}(\mytensor{A}+\mytensor{B}) &= \lambda_{k}\vec{\Phi}_{k}  \label{eq:26}\ ,
\end{align}
where $\vec{\Phi}_k = (\phi_{1k},\dots,\phi_{Nk})$ and $\vec{\Psi}_{k}=(\psi_{1k},\dots,\psi_{Nk})$. Eliminating
$\vec{\Psi}_{k}$ gives us the decoupled equation,
\begin{equation}
  \label{eq:27}
  \vec{\Phi}_{k}(\mytensor{A}-\mytensor{B})(\mytensor{A}+\mytensor{B}) = \lambda_{k}^{2}\vec{\Phi}_{k}\ .
\end{equation}
Solving the eigensystem given by~\cref{eq:27} gives the eigenvalues $\{\lambda_{i}\}$ in
the Hamiltonian~\eqref{eq:21}.

We can find the ground state energy $E_{\mathrm{g}}$ by taking trace
of~\cref{eq:17,eq:21}~\cite[Appendix A]{lieb_1961_twosoluble}. From~\cref{eq:17}, we have
\begin{equation}
  \label{eq:28}
  \Tr(H) = 2^{N-1}\sum_{i}A_{ii} -2^{N} N\Gamma
\end{equation}
and from~\cref{eq:21} we have,
\begin{equation}
  \label{eq:29}
  \Tr(H) = 2^{N-1}\sum_{k}\lambda_{k} + 2^{N}E_{\mathrm{g}}\ .
\end{equation}
As the trace is invariant under a canonical transform, we get
\begin{equation}
  \label{eq:30}
  E_{\mathrm{g}} = -N\Gamma + \frac{1}{2}(\sum_i A_{ii}- \sum_{k} {\lambda_{k}}) = -\frac{1}{2}\sum_k \lambda_k\ .
\end{equation}

If we require that the $\vec{\Phi}_{k}$ are orthonormal, then the transformation given by
~\cref{eq:22} is a canonical transformation. Solving~\cref{eq:25} gives the
corresponding $\vec{\Psi}_{k}$.

Finding matrices $\Phi$ and $\Psi$ (and hence $g_{ik}$ and $h_{ik}$) gives the forward transform
connecting the undiagonalized fermions to the diagonalized fermions. The inverse transform
can be defined as
\begin{align}
  \label{eq:31}
  \begin{split}
    a_{i}^{\dagger} &= \sum_k \bar{g}_{ik}\eta^{\dagger}_{k} + \sum_k \bar{h}_{ik} \eta_{k} \\
    a_{i} &= \sum_k \bar{g}_{ik}\eta_{k} + \sum_k \bar{h}_{ik} \eta^{\dagger}_{k}\ ,
  \end{split}
\end{align}
such that
\begin{align}
  \label{eq:32}
  \begin{split}
    a_{i}^{\dagger} + a_i &= \sum_k \bar{\phi}_{ik} \left(\eta^{\dagger}_{k} + \eta_{k} \right) \\
    a_{i}^{\dagger}  - a_{i} &= \sum_k \bar{\psi}_{ik} \left( \eta^{\dagger}_{k} - \eta_{k} \right) \ ,
  \end{split}
\end{align}
where $\phibar_{ik} = \bar{g}_{ik} + \bar{h}_{ik}$ and
$\psibar_{ik}=\bar{g}_{ik}-\bar{h}_{ik}$. Since these transforms are canonical,
$\Phibar = \Phi^{\mathrm{T}}$ and $\Psibar=\Psi^{\mathrm{T}}$.

\section{Fermionic domain-wall states}
\label{note:2}

The eigenstates of the Hamiltonian can be rewritten as many-fermion states. For example,
$\ket{0}$ denotes the vacuum which is the ground state of the Hamiltonian,
$\ket{a\ b\ c}=\etad_{a}\etad_{b}\etad_{c}\ket{0}$ is a three-fermion state with energy
$E_{\mathrm{g}}+\lambda_{a}+\lambda_{b}+\lambda_{c}$ and $\ket{\gamma}=\etad_{\{\gamma\}}\ket{0}$ is another state with
$|\gamma|$ fermions. In this notation $\ket{a\ b\ c}$ means a state with a single fermion in
each of the positions $a,b$, and $c$. What do these states look like in the computational
basis?  We can gain some intuition by considering the special case with zero transverse
field.

For $\Gamma=0$,~\cref{eq:27} becomes:
\begin{equation}
  \label{eq:33}
  \vec{\Phi}_{k} \begin{pmatrix}
    0 &       &       &        &        \\
      & 4J_{1}^{2} &       &        &        \\
      &       & 4J_{2}^{2} &        &        \\
      &       &       & \ddots &        \\
      &       &       &        &  4J_{N-1}^{2}
    \end{pmatrix}
  = \lambda_{k}^{2}\vec{\Phi}_{k}\ .
\end{equation}
For simplicity, assume $J_{1}\leq J_{2}\leq \ldots \leq J_{N-1}$.  This immediately yields the
eigenvalues $\lambda_{1}=0$, $\lambda_{2}=2J_{1}$, $\lambda_{3}=2J_{2}$, $\ldots$,
$\lambda_{N}=2J_{N-1}$ and eigenvector
\begin{align}
  \Phi &= \begin{pmatrix}
    1 &       &       &        &        \\
      & 1 &       &        &        \\
      &       & 1 &        &        \\
      &       &       & \ddots &        \\
      &       &       &        &  1
    \end{pmatrix}\ . \label{eq:34}
\end{align}
Using~\cref{eq:26} we find:
\begin{align}
  \Psi &= \begin{pmatrix}
     0 &   0 &   0 &  \ldots   &   1     \\
    -1 &   0 &   0 &  \ldots   &   0     \\
     0 &  -1 &   0 &  \ldots   &   0     \\
     0  &    &  \ddots  &      &  0      \\
     0  &  0 &  \ldots   &   -1 &  0
    \end{pmatrix}\ . \label{eq:35}
\end{align}
If the $J_{i}$ are not ordered the result remains the same up to a permutation of rows
of $\Phi$ and $\Psi$, where the $\lambda_{i}$ are still arranged in ascending order.

Now consider the operator
$\sigz_{i}\sigz_{i+1}=(a^{\dagger}_{i}-a_{i})(a^{\dagger}_{i+1}+a_{i+1})$ in terms of the diagonalized
fermionic operators. Using~\cref{eq:32,eq:34,eq:35} gives
\begin{align}
  \label{eq:36}
    \sigz_{i}\sigz_{i+1}
    &= \sum_{kk'} \bar{\psi}_{ki}\bar{\phi}_{i+1,k'}(\etad_{k}-\eta_{k})(\etad_{k'}+\eta_{k'}) \notag \\
    &= \sum_{kk'} (-\delta_{k,i+1})(\delta_{k',i+1})(\etad_{k}-\eta_{k})(\etad_{k'}+\eta_{k'}) \notag \\
    &= 1-2\etad_{i+1}\eta_{i+1}\ .
\end{align}
Thus, the many-fermion states are also the eigenstates of the operators
$\sigz_{i}\sigz_{i+1}$ with eigenvalue $-1$ if there is a fermion occupying level $i+1$,
and eigenvalue $1$ otherwise. In the spin picture, eigenvalue $+1$ denotes a satisfied
coupling and $-1$ denotes an unsatisfied coupling. Thus, the presence of a fermion in
level $i+1$ can be thought as a domain wall in coupling $i$ of the ferromagnetic
chain. The fermionic states $\ket{0}$ and $\ket{1}$ satisfy all the couplings and hence
are linear combinations of the all-$0$ and all-$1$ states. This is reflective of the
$\mathbb{Z}_{2}$ symmetry inherent in the Ising Hamiltonian, which also manifest itself as
$\lambda_{1}=0$. Thus, for given state $\ket{a}$ with energy $E_{a}=E_{\mathrm{g}}+\lambda_{a}$, the state
$\ket{a\ 1}$ has the same energy $E_{\mathrm{g}}+\lambda_{a}+\lambda_{1}=E_{a}$. Additionally,
$(1-2\etad_{i+1}\eta_{i+1})\ket{a}=(1-2\etad_{i+1}\eta_{i+1})\ket{a\ 1}\ \forall\ i\geq1$ since the
fermion occupying the state $\ket{1}$ is not affected by this operation. Ergo, the states
$\ket{2}$ and $\ket{2\ 1}$ corresponds to a domain wall at the location of the weakest
coupling, $\ket{3}$ and $\ket{3\ 1}$ corresponds to a state with a domain wall at the
second weakest coupling, etc. Once the couplers are arranged in alternating sectors such
that $J_{1}=J_{2}=\ldots=W_{1}>J_{n+1}=J_{n+2}=\ldots=W_{2}<\ldots$, the domain walls first occur in the
light sectors, followed by the heavy sectors, etc.

\section{Why states that differ by a single fermionic excitation have the largest matrix elements}
\label{note:3}

In order to determine whether thermal transitions between states can occur, we need to
calculate the matrix element of $\sigz$ between the two states, where we assume that the
dominant interaction term between the system and environment is given by a pure dephasing
interaction of the form
$H_{\mathrm{SB}}=\sum_{i} \sigz_{i} \otimes
B_{i}$~\cite{lanting_2010_cotunnelingpairs,boixo_2016_computationalmultiqubit,albash_2015_consistencytests,albash_2015_reexaminationevidence,albash_2015_consistencytests}.
We we would like to estimate the transition matrix element
$\braket{\gamma_{1}|\sigz_{i}|\gamma_{2}}$ between states $\ket{\gamma_{1}}$ and
$\ket{\gamma_{2}}$. We shall show that these transitions are most prominent when the states
differ by only a single fermion.

In terms of the fermionic operators, the operator $\sigz$ can be written as
\begin{equation}
  \label{eq:37}
  \sigz_{i} = \left[ \prod_{j=1}^{i-1}(a_{j}^{\dagger}+a_{j})(a_{j}^{\dagger}-a_{j})\right](a^{\dagger}_{i}+a_{i})
\end{equation}
and it can in turn be written in term of the $\{\eta_k\}$ and $\{\eta_k^{\dagger}\}$ operators using
~\cref{eq:32}.

Now consider the matrix element between the vacuum state $\ket{0}$ and the state
$\ket{\gamma}$. For the operator $\sigz_{1}$ we find that
\begin{align}
  \label{eq:38}
  \begin{split}
    \braket{\gamma|\sigz_{1}|0} &= \braket{\gamma | a^{\dagger}_{1} + a_{1} | 0 } \\
    &= \braket{\gamma | \sum_{k} \phibar_{1k}(\eta^{\dagger}_{k}+\eta_{k})|0 } \\
    &= \sum_{k} \phibar_{1k} \braket{\gamma|k}.
  \end{split}
\end{align}
This can only be non-zero if $\ket{\gamma}$ is a single fermion state.  Thus, the operator
$\sigz_{1}$ connects the vacuum to single-fermion states.

For $\sigz_{2}$, we have
\begin{align}
  \label{eq:39}
  \begin{split}
    &\braket{\gamma|\sigz_{2}|0} = \braket{\gamma |(a^{\dagger}_{1} + a_{1})(a^{\dagger}_{1} - a_{1}) (a^{\dagger}_{2} + a_{2}) | 0 } \\
    &= \sum_{k,l,m}\phibar_{1k}\psibar_{1l}\phibar_{2m} \braket{\gamma |
      (\eta^{\dagger}_{k}+\eta_{k})(\eta^{\dagger}_{l}+\eta_{l})(\eta^{\dagger}_{m}+\eta_{m})|0 }.
  \end{split}
\end{align}
For the above term to be non-zero, the state $\ket{\gamma}$ must be either a three-fermion or a
single-fermion state. The one-fermion case can be computed in a similar manner to the
previous case involving $\sigma_1^z$, so we focus on the case when
$\ket{\gamma}=\ket{a\ b\ c}$ is a three-fermion state. We have
\begin{align}
  \label{eq:40}
  \begin{split}
    &\braket{a\  b\ c|\sigz_{2}|0} \\
    &= \sum_{klm}\phibar_{1k}\psibar_{1l}\phibar_{2m}\braket{a\
      b\ c|\eta_{k}^{\dagger}\eta_{l}^{\dagger}\eta_{m}^{\dagger}|0} \\
    &= \sum_{klm}\phibar_{1k}\psibar_{1l}\phibar_{2m}\braket{a\
      b\ c|k\ l\ m} \\
    &= \det{\begin{vmatrix}
        \phibar_{1a} & \psibar_{1a} & \phibar_{2a} \\
        \phibar_{1b} & \psibar_{1b} & \phibar_{2b} \\
        \phibar_{1c} & \psibar_{1c} & \phibar_{2c}
      \end{vmatrix}}\ .
  \end{split}
\end{align}
We find numerically for our problems that the matrix element associated with the
three-fermion states is much smaller than that for single-fermion states. For example, for
a chain of length $N=176$ with parameters $n=5$, $W_{1}=1$ and $W_2=0.5$, we find at
$s=s^{\ast}$ that $\braket{3|\sigz_{1}|0}=0.12$ and
$\braket{2\ 3\ 4|\sigz_{2}|0} = 1.2 \times {10}^{-8}$.

Similarly, the matrix element involving $\sigz_{3}$ requires $\ket{\gamma}$ to be a state with
$1$, $3$ or $5$ fermions, etc. We find that the excitation to the $5$-fermion states will
be even smaller than that to the $3$-fermion states, since they contain terms that are the
product of five terms of the type $\phibar\psibar\phibar\psibar\phibar$.  Thus, our
numerical results indicate that the vacuum state couples predominantly to single-fermion
states.

The above analysis can be generalized. Let us consider the matrix
element $\theta_{ij}= \braket{0|\sigz_{i}|j}$. Let
$\mytensor{\Theta}=[\theta_{ij}]$ and
$\mytensor{\Psibar}=[\psibar_{ij}]$, and consider
$\mytensor{\tildeG} = \mytensor{\Theta}\mytensor{\Psibar}^{\mathrm{T}}$. We have:
\begin{align}
    \tildeG_{ij} &= \sum_{m} \theta_{im}{\left(\mytensor{\Psibar}^{\mathrm{T}}\right)}_{mj} \label{eq:41}\\
    &= \sum_{m} \psibar_{jm} \braket{0|\sigz_{i}|m} \label{eq:42} \\
    &= \braket{0|\sigz_{i}\left(\sum_{m}\psibar_{jm}(\eta_{m}^{\dagger}-\eta_{m})\right)|0} \label{eq:43} \\
    &= \braket{0|\sigz_{i}(a_{j}^{\dagger}-a_{j})|0} \label{eq:44} \\
    &= \braket{0|A_{1}B_{1} \ldots A_{i-1}B_{i-1}A_{i}B_{j}|0}\ ,  \label{eq:45}
\end{align}
where we have defined $A_{i}=(a_{i}^{\dagger}+a_{i})$ and $B_{i}=(a_{i}^{\dagger}-a_{i})$.  Since the
operators $A_i$ and $B_j$ in~\cref{eq:45} anticommute when they have different
indices, we can simplify the expression using Wick's
theorem~\cite{wick_1950_evaluationcollision,lieb_1961_twosoluble}. For a set of anticommutating operators
$\{O_{1},O_{2},\ldots,O_{2n}\}$, Wick's theorem states that
\begin{multline}
  \label{eq:46}
  \braket{0|O_{1}O_{2}\ldots O_{2n}|0} \\
  = \sum_{\text{possible pairings}}(-1)^{p} \prod_{\text{all pairs}}\braket{0|O_{i_{1}}O_{i_{2}}|0}\ .
\end{multline}
where the sum is over all possible pairings of the operators
$\{O_{1},O_{2},\ldots,O_{2n}\}$, the product is over the two-point expectation value of all
pairs, and $(-1)^{p}$ is the sign of the permutation that is required to bring the paired
terms next to each other. For an odd number of operators, the expectation value
vanishes. Applying the theorem to~\cref{eq:45}, we can make the following
simplifications:
\begin{equation}
  \label{eq:47}
  \begin{aligned}
    &\braket{0|A_{i}A_{j}|0}=\phantom{-}  \sum_{k}\phibar_{ik}\phibar_{jk} = \delta_{ij}\ ,\\
    &\braket{0|B_{i}B_{j}|0}= -\sum_{k}\psibar_{ik}\psibar_{jk} = -\delta_{ij}\ ,\\
    &\braket{0|A_{i}B_{j}|0}=\phantom{-} \sum_{k}\phibar_{ik}\psibar_{jk}
    =\left(\mytensor{\Phibar}\mytensor{\Psibar}^{\mathrm{T}}\right)_{ij} \equiv G_{ij}\ .
  \end{aligned}
\end{equation}
If $i<j$, all the terms in~\cref{eq:45} will have different indices. The non-zero
terms are pairs of the form $\braket{A_{k}B_{k'}}$. An obvious pairing is
$\braket{A_{1}B_{1}}\ldots\braket{A_{i}B_{j}} = G_{11}G_{22}\ldots G_{ij}$ and all other
permutations can be obtained by keeping the $A$'s fixed and permuting the $B$'s around
them. The signature of the permutation will be the signature of the permutations of
$B$'s. The sum over permutations $p$ is then given by
\begin{equation}
  \label{eq:48}
  \begin{aligned}
    \tildeG_{ij} &=
    \sum_{p}(-1)^{p}G_{1p_{1}}G_{2p_{2}} \ldots G_{ip_{i}} \\
    &= \det{
      \begin{vmatrix}
        G_{11} & G_{12} & \ldots & G_{1j} \\
        G_{21} & G_{22} & \ldots & G_{2j} \\
        \vdots      &        &   & \\
        G_{i1} & G_{i2} & \ldots & G_{ij}
      \end{vmatrix}}\ .
  \end{aligned}
\end{equation}

If $i \geq j$, we need to further simplify~\cref{eq:45} so that it only contains
anticommuting terms;
\begin{align}
  \label{eq:49}
  \begin{split}
    \braket{A_{1}B_{1} \ldots A_{j} B_{j} \ldots A_{i}B_{j}} &=
    -\braket{A_{1}B_{1}\ldots A_{j}B_{j}B_{j}\ldots A_{i}} \\
    &= -\braket{A_{1}B_{1}\ldots A_{j}(-1)\ldots A_{i}} \\
    &= \braket{A_{1}B_{1}\ldots A_{j}\ldots A_{i-1}B_{i-1}A_{i}}\ .
  \end{split}
\end{align}
In Wick's expansion of this equation, any possible permutation will contain a pair of form
$\braket{A_{k}A_{k'}}=0$. Thus, $\tilde{G}_{ij}=0$ for $i>j$.

This gives us a method to calculate the matrix elements of $\mytensor{\tildeG}$, and we can
in turn compute the desired transition element matrix via
\begin{equation}
  \label{eq:50}
  \mytensor{\Theta} = \mytensor{\tildeG}\mytensor{\Psibar}\ .
\end{equation}

Matrix element between arbitrary states can be computed in a similar fashion.  For
example:
\begin{align}
  \label{eq:51}
  \begin{split}
    \braket{\gamma_{1}|\sigz_{i}|\gamma_{2}} &= \braket{\gamma_{1}|A_{1}B_{1}\ldots A_{i-1}B_{i-1}A_{i}|\gamma_{2}} \\
    &= \braket{\gamma_{1}|A_{1}B_{1}\ldots A_{i-1}B_{i-1}A_{i}\eta^{\dagger}_{\{\gamma_{2}\}}|0} .
  \end{split}
\end{align}
Since $A_{i}=\sum_{j}\phibar_{ij}(\etad_{j}+\eta_{j})$ and
$B_{i}=\sum_{j}\psibar_{ij}(\etad_{j}-\eta_{j})$ we have
\begin{align}
  \label{eq:52}
  \begin{split}
    \{A_{i},\etad_{j}\} &= \phantom{-}\phibar_{ij}\ , \\
    \{B_{i},\etad_{j}\} &= -\psibar_{ij}\ .
  \end{split}
\end{align}
We can therefore anticommute the creation operator of $\etad_{\{\gamma_{2}\}}$ from the left
hand side to the right hand side.  We find that the additional terms appearing because of
the anticommutation relation in~\cref{eq:52} are small relative to the term
proportional to
$\braket{\gamma_{1}|\etad_{\{\gamma_{2}\}}A_{1}B_{1}\ldots A_{i-1}B_{i-1}A_{i}|0}$, so we focus only on
this term:
\begin{align}
  \label{eq:53}
    & \braket{\gamma_{1}|\sigz_{i}|\gamma_{2}} \\
    &\quad \sim (-1)^{(2i-1)|\gamma_{2}|}
    \braket{\gamma_{1}|\etad_{\{\gamma_{2}\}}A_{1}B_{1}\ldots A_{i-1}B_{i-1}A_{i}|0} \notag \\
    &\quad = (-1)^{(2i-1)|\gamma_{2}|}\braket{\gamma_{1}-\gamma_{2}|A_{1}B_{1}\ldots A_{i-1}B_{i-1}A_{i}|0} \notag\\
    &\quad = (-1)^{(2i-1)|\gamma_{2}|}\braket{\gamma_{1}-\gamma_{2}|\sigz_{i}|0}\ , \notag
\end{align}
where $(-1)^{(2i-1)|\gamma_{2}|}$ is an overall phase term associated with anticommuting the
set $\etad_{\{\gamma_{2}\}}$ to the left and
$\ket{\gamma_{1}-\gamma_{2}}=\eta_{\{\gamma_{2}\}}\ket{\gamma_{1}}$. For this term to be non zero, the state
$\ket{\gamma_{1}}$ should contain all the fermions found in the state $\ket{\gamma_{2}}$. The
remaining quantity $\braket{\gamma_{1}-\gamma_{2}|\sigz_{i}|0}$ is a matrix element connecting the
vacuum, which we have already argued is largest when the state
$\ket{\gamma_{1}-\gamma_{2}}$ is a single-fermion state. Therefore, the matrix elements connecting
$\ket{\gamma_{1}}$ and $\ket{\gamma_{2}}$ are largest when the two states differ only by one
fermion.

From the above analysis of the coupling matrix elements, we find that the ground state
couples principally with the single-fermion states, and the single-fermion states couple
principally to two-fermion states and so on. The transition energies are the
single-fermion energies $\{\lambda_{i}\}$ found by solving the eigensystem of~\cref{eq:27}.
\section{Exponential dependence of the success probability on $k^{\ast}$}
\label{note:4}

In this section we provide a more detailed argument than given in the main text for why
the success probability exhibits an exponential dependence on $k^{\ast}$.  Rather than
providing a counting argument based on the thermal density of states $d$, as in
\cref{eq:7}, we consider the transition rates appearing in the Pauli
master equation. Our argument provides a justification for why, if it were numerically
feasible, we would expect the master equation simulations to reproduce the exponential
dependence on $k^*$ seen in our empirical results.

Given the form of the Pauli master equation [~\cref{eq:8}],
$\dot{p}_{a} = \sum_{b \neq a} \gamma(\omega_{ba}) M_{ab} p_{b} - \left( \sum_{b \neq a} \gamma(\omega_{ab}) M_{ba} \right
) p_{a}$, where [~\cref{eq:9}]
$M_{ab} = \sum_{\alpha} |\braket{a|A_{\alpha}|b}|^{2}$, we expect the success probability to be
inversely related to the overall excitation rate. Let $M_{0b}\equiv M_{b}$ be the matrix
element involving a transition between states via the creation or annihilation of a
fermion with energy $\lambda_b$, where $\lambda_{b}\leq\lambda_{k^{\ast}}$. Let the number of such connected
states be $\#_{b} = {k^* \choose b}$ and let $M_{\min}$ ($M_{\max}$) be the minimum
(maximum) matrix element within the set $\{M_{1},M_{2},\ldots,M_{k^{\ast}}\}$. The total
transition rate $\tau$ is can be estimated as
\begin{equation}
  \label{eq:54}
  \tau \sim M_{1}\times\#_{1} + M_{2}\times\#_{2} + \ldots + M_{k^{\ast}}\times\#_{k^{\ast}}\ ,
\end{equation}
so that
\begin{equation}
  \label{eq:55}
  M_{\min}  2^{k^{\ast}} \lesssim \tau \lesssim M_{\max}  2^{k^{\ast}}\ ,
\end{equation}
where we used $\sum_{b=0}^{k^*} \#_{b} = 2^{k^*}$.
This bound is meaningful if the matrix elements do not differ by orders of
magnitude. Indeed, we found numerically that the matrix elements are within an order of
magnitude of each other. For example, for the chain with parameters $N=176$, $n=5$,
$W_{1}=1$ and $W_{2}=0.5$, we found that at the critical point with $k^{\ast}=18$, the
largest matrix element is $M_{\max}=M_{1}=43.76$ and the smallest matrix element is
$M_{\min}=M_{18}=0.99$. The difference in these values is small compared to the number of
fermionic states, which is $2^{18}$. Thus, $\tau=\Omega(2^{k^{\ast}})$ and we expect the success
probability to be inversely proportional to $2^{k^{\ast}}$.

\begin{figure}[t]
  \centering
  \includegraphics[width=\columnwidth,height=0.75\columnwidth]{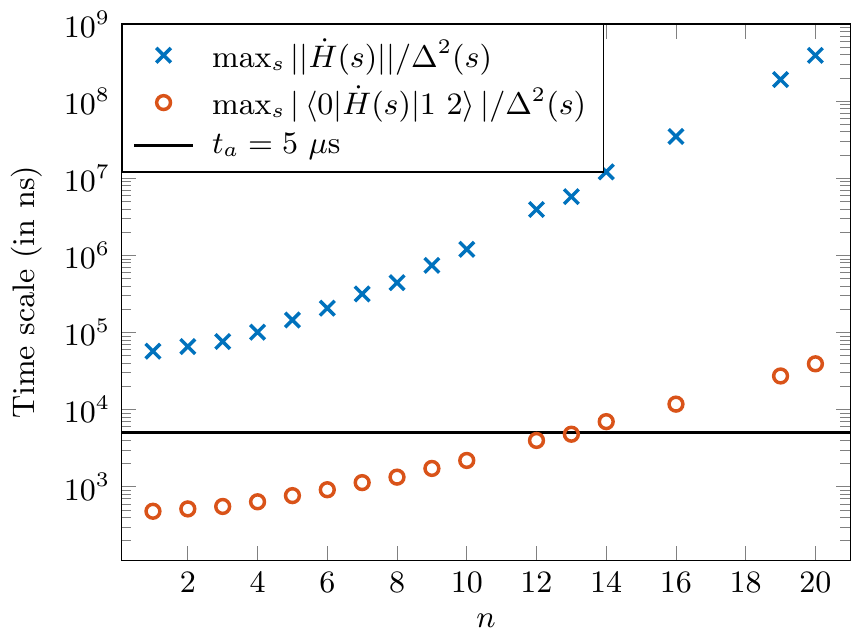}
  \caption{\textbf{Test of the adiabatic condition}. The
    solid line is the annealing time used in our experiments. Symbols represent the
    quantity appearing in two versions of the adiabatic condition [for ASC parameters
    $(1,0.5,175)$] that should be smaller than the annealing time in order for the
    adiabatic condition to hold.}
  \label{fig:7}
\end{figure}

\section{Test of the adiabatic condition}
\label{note:5}

Here we test the validity conditions assumed for the derivation of the adiabatic Markovian
master equation~\cite{albash_2012_quantumadiabatic}, and in particular that the adiabatic
approximation is satisfied. To this end we test the `folklore' adiabatic condition
$t_f \gg \max_s |\langle g|\dot{H}(s)|e\rangle|/\Delta^2(s)$, where $g$ and $e$ are the ground and first
(relevant) excited states, respectively, for ASC parameters $(1,0.5,175)$. The result is
shown in Supplementary Fig.~\ref{fig:7} (red circles), and it can be seen that the
condition is satisfied for $n\leq 14$. The relevant first excited state is the two-fermion
state $|1 \ 2\rangle$. Also shown is the more conservative condition given in terms of the
operator norm (blue crosses). The adiabatic condition with the operator norm is not
satisfied for any value of $n$, but it has recently been shown that this condition, which
involves an extensively growing operator norm, must be replaced by a condition involving
local operators~\cite{bachmann_2017_adiabatictheorem}.
\begin{figure*}
  \centering \includegraphics[width=\textwidth]{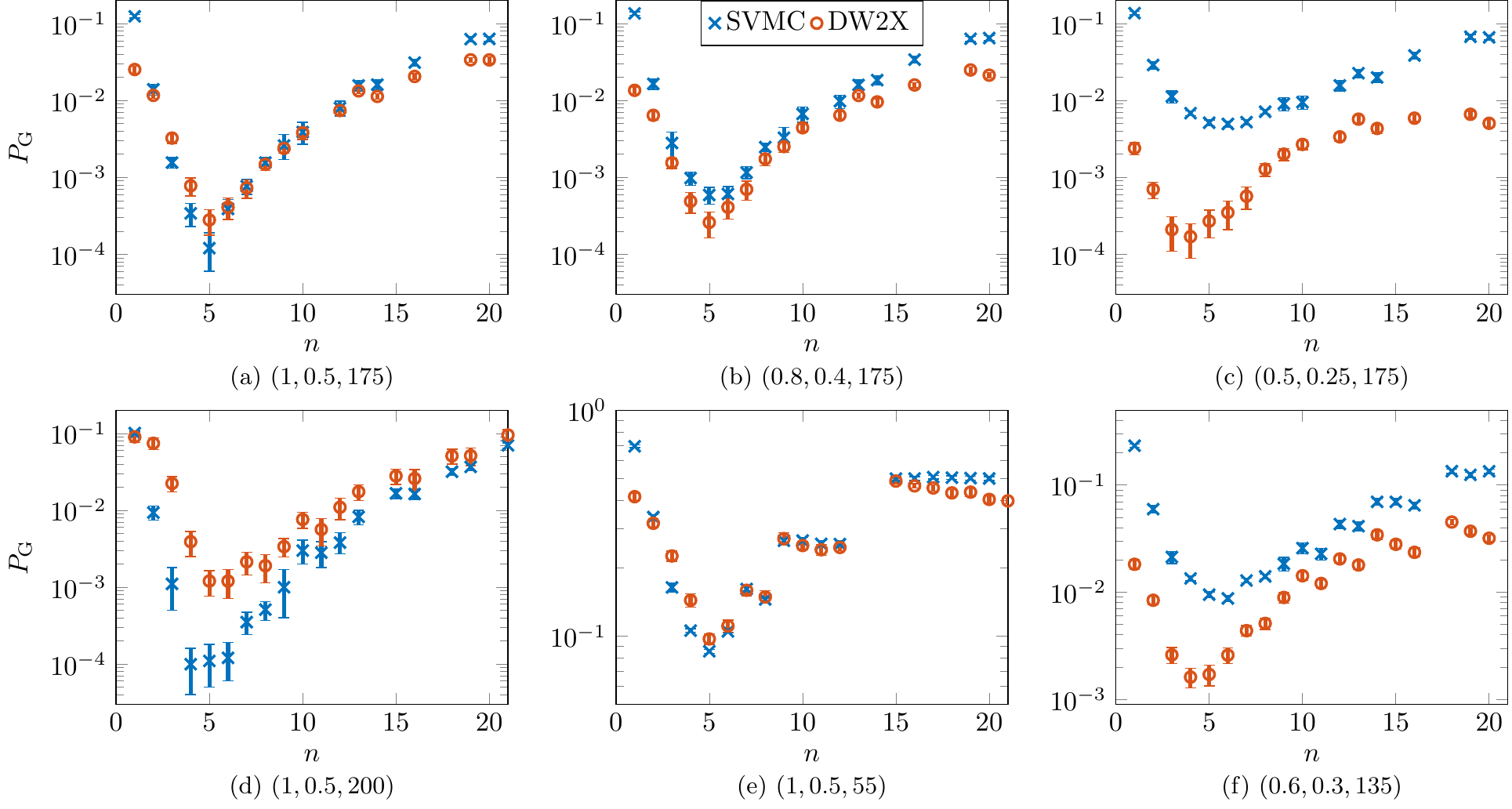}
  \caption{\textbf{Comparison of the SMVC model to the empirical DW2X results}. The error
    bars everywhere indicate $95\%$ confidence intervals calculated using a binomial
    bootstrap over the different runs of the simulation. (a) The SVMC parameters were
    optimized to match the empirical DW2X success probability results for the chain with
    parameters $(W_1,W_2,N)=(1,0.5,175)$. The optimal values found are:
    $N_s=120\times10^{3}, \beta=0.75 \text{ (GHz)}^{-1}, \sigma=0.05$ [compare to the DW2X's
    $t_f=5\ \mu\text{s}, \beta=0.637 \text{ (GHz)}^{-1}, \sigma\sim0.03$].  (b) With the same optimal SVMC
    parameter values, but with chain parameters $(0.8,0.4,175)$, the SVMC model predicts
    increased success probability, in contrast to the empirical results. The same trend
    continues but is more pronounced in (c), with additionally the position of the minimum
    shifting to the wrong location ($n=7$ vs $n^*=4$). Panel (d) shows that
    increasing the chain length causes a large deviation in the SVMC results [compare to
    panel (a)], and also shifts the location of the minimum to the wrong value, but (e)
    shows showing that reducing the chain length does not degrade the agreement much. (f)
    Results for another chain parameter set, exhibiting a similar discrepancy to that seen
    in (c).}
  \label{fig:8}
\end{figure*}
\begin{table}
  \centering
  \begin{tabular}{ccccc}
    Chain parameters        	& DW2X	& $k^*$	& ME	& SVMC   \\
    \hline
    $(1.0,0.50,175)$  		& 5,6	& 5    	& 5,6	& 5      \\
    $(0.5,0.25,175)$  		& 3,4,5	& 4	& 4,5	& 5,6,7  \\
    $(0.8,0.40,175)$  		& 5	& 5     & 5	& 5,6    \\
    $(1.0,0.50, 200)$  		& 5,6	& 5    	& 5,6	& 4,5,6  \\
    $(1.0,0.50, \phantom{0}55)$ & 5,6	& 5    	& 5,6,8	& 5      \\
    $(0.6,0.30,135)$  		& 4,5	& 4,5 	& 5,6	& 5,6
  \end{tabular}
  \caption{Locations $n^*$ of the DW2X success probability minima vs those found
    by the fermionic model based on the peak of $k^*$, the master equation model (ME), and
    the SVMC model. When the location of the minimum is ambiguous within our $95\%$
    confidence interval we list all values of $n^*$ that overlap to within one $\sigma$. The
    best agreement is obtained by the master equation model.}
  \label{tab:2}
\end{table}
\section{Comparison to the classical SVMC model}
\label{note:6}
The spin vector Monte Carlo (SVMC) model~\cite{shin_2014_howquantum} was proposed as a purely
classical model of the D-Wave processors in response to earlier work that ruled out
simulated annealing \cite{boixo_2013_experimentalsignature} and other work that established a strong correlation
between D-Wave ground state success probability data and simulated quantum annealing
\cite{boixo_2014_evidencequantum}. The SVMC model was found to not always correlate well with D-Wave empirical
data; for example, deviations were observed for the SVMC model in the case of ground
state degeneracy breaking~\cite{albash_2015_consistencytests}, excited state
distributions~\cite{albash_2015_reexaminationevidence}, quantum annealing correction
experiments~\cite{pudenz_2015_quantumannealing} and the dependence of success probability on
temperature~\cite{boixo_2016_computationalmultiqubit}. But it has generally been successful in
predicting the success probability distributions of D-Wave experiments. The SVMC model
thus provides a sensitive test for whether anything other than classical effects are at
play in a fixed-temperature measurement of the ground state success probability.

In the SVMC model each qubit is replaced by a classical $O(2)$ rotor parametrized by one
continuous angle via $\sigma^x_i\mapsto \sin\theta_i$ and $\sigma^z_i\mapsto \cos\theta_i$, so that the
Hamiltonian~[\cref{eq:1}] becomes
\begin{equation}
  \label{eq:56}
  E(s) = -A(s) \sum_{i} \sin \theta_{i} + B(s) \left(-\sum_{i}J_{i}\cos \theta_{i}\cos\theta_{i+1} \right)\ .
\end{equation}
In addition, the angles $\theta_{i}$ undergo Metropolis updates every discrete time-step of
size $1/N_s$, where $N_s$ is the number of sweeps, at a fixed inverse temperature $\beta$. \\

Angle updates are performed as follows. We first divide the dimensionless time $s$ into
steps of size $\delta s = 1/N_s$, where $N_s$ is the number of sweeps performed during the
course of one run, and initialize to the state $\theta_i(s=0)=\pi/2$ for all spins $i$.  At each
such time step, we pick a random permutation of the set $\{1,2,\ldots,N\}$ where $N$ is the
number of qubits in the chain. One by one, we select a random angle for each qubit in this
permuted list, changing it from $\theta_{i}$ to $\tilde{\theta}_{i}$ where
$\tilde{\theta}_{i}$ is picked randomly from $[0,\pi]$. We calculate the energy change
$\Delta E$ for this move and the new angle is accepted with probability
\begin{equation}
  \label{eq:57}
  p_{i} = \min \left[ 1, \exp (-\beta \Delta E) \right]\ ,
\end{equation}
where $\beta$ is the inverse annealing temperature for the SVMC algorithm. The final state is
obtained by projecting the $O(2)$ rotor spins to the computational state at $s=1$, setting
spin $i$ to be $1$ ($-1$) if $\cos\theta_{i}>0$ ($<0$). We repeat this process many times in
order to estimate the success probability of the algorithm.

In order to realistically emulate the D-Wave processor, we added random Gaussian noise
$\mathcal{N}(0,\sigma^2)$ to each $J_i$~\cite{shin_2014_commentdistinguishing}. The SVMC model then has three free
parameters: $\{N_s,\beta,\sigma\}$, which we used to calibrate it against the DW2X data. Toward
this end we used the chain with parameters $(1,0.5,175)$ and performed an extensive search
in the $\{N_s,\beta,\sigma\}$ parameter space. As shown in Supplementary Fig.~\ref{fig:8}(a), we
obtain a close match for $N_s=120\times10^{3}$, $\beta=0.75 \text{ (GHz)}^{-1}$ and
$\sigma=0.05$. This is the best fit we found for this particular chain. In general, we found
that the SVMC parameters can be tuned to reproduce the location of the minimum in success
probability for any sector size. We were also able to tune the parameters such that the
minimum disappears completely and have the success probability increase or decrease
monotonically.  We were not able to find parameters that give rise to an inverted curve,
i.e., a maximum in the success probability. Most of these features can be seen by tuning
$\beta$ and keeping the other parameters fixed.
\begin{figure}
  \includegraphics[width=0.9\columnwidth,height=0.675\columnwidth]
  {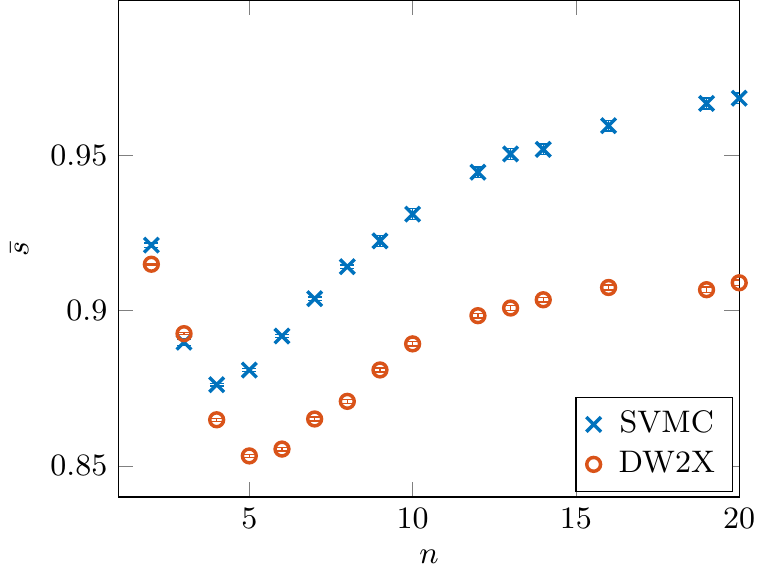}
  \caption{\textbf{Spin boundary correlation function computed using the SVMC model.}  The
    spin boundary correlation function for the same chain and SVMC parameters as in
    Supplementary Fig.~\ref{fig:8}(a). The SVMC model does not correctly capture the
    empirical results despite providing a close match to the success probability in
    Supplementary Fig.~\ref{fig:8}(a).}
  \label{fig:9}
\end{figure}

To avoid fine-tuning, we next used the same parameters to compute the success probability
of the SVMC model for other chains. As shown in Supplementary Figs.~\ref{fig:8}(b) and
\ref{fig:8}(c), the SVMC model has the wrong trend with decreasing $(W_1,W_2)$: it
exhibits a higher success probability as the coupling energy scale is lowered. The same
happens with increased chain length [Supplementary Fig.~\ref{fig:8}(d)], though to a
lesser degree with decreased chain length [Supplementary Fig.~\ref{fig:8}(e)]. Moreover,
as summarized in Table~\ref{tab:2}, it does not agree as well with the location of the
minimum of the success probability as the other models.  We {emphasize} that while we
performed an extensive search, we cannot rule out the possibility of another set of
parameters (or the inclusion of other parameters) that allow SVMC to reproduce the DW2X
results for all chain
lengths and energy scales. \\

As a further test we also considered the spin boundary correlation, defined as the
sum of spin correlations over all boundary qubits in the chain, where the boundary qubits
are the qubits at the right edge ($r$) of the heavy sector and left edge ($l$) of the
light sector:
\begin{equation}
  \label{eq:58}
  \bar{s} = \frac{1}{|Q|} \sum_{Q}s_{l}s_{r}\ ,
\end{equation}
where $Q$ is the set of boundary qubits and $s_l,s_r\in\{0,1\}$. Thus $\bar{s}=1$ represents
perfect alignment (a ground state), while $\bar{s}<1$ represents the occurrence of an
excited state due to misalignment of the different sectors. Supplementary Fig.~\ref{fig:9}
shows the results for the same set of optimized parameters that provided strong agreement
with the ground state success probability in Supplementary Fig.~\ref{fig:8}(a). It can be
seen that the SVMC model predicts the wrong location for the minimum of $\bar{s}$ and
rises too fast. Unfortunately, master equation simulations for $\bar{s}$ are numerically
prohibitive, so we cannot assess whether this discrepancy of the SVMC model is fixed by a
quantum model.\\

\section{Results at different annealing times}
\label{note:7}

\begin{figure}[b]
  \centering
  \includegraphics[width=0.9\columnwidth,height=0.675\columnwidth]
  {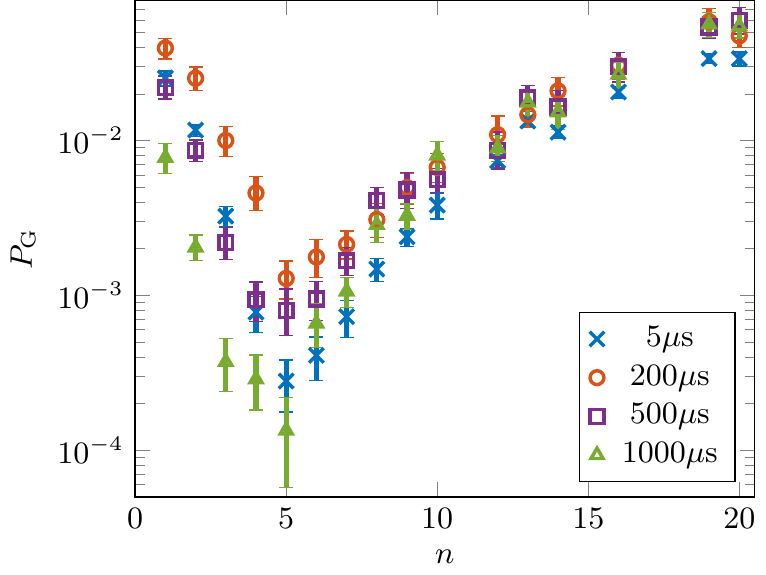}
  \caption{\textbf{Dependence of the success probability on the annealing times.} We show
    the results for the ASC with parameters $(1.0,0.5,175)$. The error bars everywhere
    indicate $95\%$ confidence intervals calculated using a bootstrap over different
    gauges and embeddings. The location of the minimum is unchanged as the annealing time
    is varied.}
  \label{fig:10}
\end{figure}

In~\ref{note:5} we discussed the validity of the ``folklore'' adiabatic
condition for the ASC problem. We expect that the adiabatic condition will also be
satisfied if we make small changes in the annealing time compared to the vertical scale of
Supplementary Fig.~\ref{fig:7}. Also, neither the gap $\Delta$ nor the number of single fermion
states $k^{\ast}$ changes with an increase in the annealing time. Hence we expect that the
qualitative nature of the success probability curve, including the location of minima,
will be independent of small changes in the annealing time. Since the total thermal
transition rate depends on the amount of time system spends near the quantum minimum gap
point $s^{\ast}$, we do expect to see changes in the value of the success probability. In
Supplementary Fig.~\ref{fig:10}, we show the change in success probability as we vary
the annealing time on the D-Wave device. As expected, the location of the minima remains
unchanged. We do find that the success probability varies depending on the annealing time
and sector size.

\end{document}